\documentclass[11pt,a4paper]{article}
\usepackage{epsfig}
\usepackage[T1]{fontenc}    
\usepackage[latin1]{inputenc} 
\usepackage{graphics}
\usepackage{graphicx}
\usepackage{psfrag}
\usepackage{pstricks,pst-coil,pst-fill,pst-plot}
\usepackage{pgf}  
\usepackage[fleqn]{amsmath}    
\usepackage{amssymb}    
\usepackage{amsfonts}   
\usepackage{verbatim}   
\usepackage{mathrsfs}   
\usepackage{dsfont}
\usepackage{euscript}
\usepackage{yfonts}
\usepackage{txfonts}
\usepackage{marvosym}
\usepackage{vmargin}        

\setmarginsrb{1.8cm}{2cm}{1.8cm}{2cm}{1cm}{1cm}{1cm}{1.6cm}
 \makeatletter
 \@addtoreset{equation}{section}
 \makeatother

\providecommand{\bysame}{\leavevmode\hbox to3em{\hrulefill}\thinspace}

\let\tend=\rightarrow

\def\qed{\hfill\rule{2mm}{2mm}}

\newcommand\beq{\begin{equation}}
\newcommand\enq{\end{equation}}
\newcommand\bem{\begin{multline}}
\newcommand\enm{\end{multline}}

\def\beqa{\begin{eqnarray}}
\def\eeqa{\end{eqnarray}}
\def\ba{\begin{array}}
\def\ea{\end{array}}

\def\det{\operatorname{det}}

\newcommand{\f}[2]{{\ensuremath{%
    \mathchoice%
    {\dfrac{#1}{#2}}
    {\dfrac{#1}{#2}}
    {\frac{#1}{#2}}
    {\frac{#1}{#2}}
}}}
\newcommand{\tf}[2]{\ensuremath{#1/#2}}
\newcommand{\pa}[1]{\ensuremath{\left(#1\right)}}
\newcommand{\paa}[1]{\ensuremath{\left\{#1\right\}}}

\newcommand{\pab}[2]{\ensuremath{\pa{\ba{c} #1 \\ #2 \ea }}}

\def\Ga{\Gamma}
\def\de{\delta}

\def\De{\Delta}
\def\eps{\epsilon}
\def\veps{\varepsilon}
\def\la{\lambda}

\def\sg{\sigma}

\def\th{\theta}

\def\om{\omega}
\def\vp{\varphi}

\newcommand{\mc}[1]{\ensuremath{\mathcal{#1}}}
\newcommand{\mf}[1]{\ensuremath{\mathfrak{#1}}}
\newcommand{\msc}[1]{\ensuremath{\mathscr{#1}}}

\newcommand{\bs}[1]{\ensuremath{\boldsymbol{#1}}}

\newcommand{\ov}[1]{\ensuremath{\overline{#1}}}
\newcommand{\wt}[1]{\ensuremath{\widetilde{#1}}}
\newcommand{\wh}[1]{\ensuremath{\widehat{#1}}}

\newcommand{\Int}[2]{\ensuremath{\int\limits_{#1}^{#2}}}

\newcommand{\sul}[2]{\ensuremath{\sum\limits_{#1}^{#2}}}
\newcommand{\pl}[2]{\ensuremath{\prod\limits_{#1}^{#2}}}

\newcommand{\R}{\ensuremath{\mathbb{R}}}
\newcommand{\Cx}{\ensuremath{\mathbb{C}}}

\newcommand{\Dp}[1]{\ensuremath{\partial_{#1}}}

\newcommand{\ex}[1]{\ensuremath{\e{e}^{#1}}}

\newcommand{\bra}[1]{ \langle \,#1\,|}
\newcommand{\ket}[1]{ |\,#1\, \rangle}

\newcommand{\dd}{\mathrm{d}}
\newcommand{\e}[1]{\ensuremath{\mathrm{#1}}}

\newcommand{\intff}[2]{\ensuremath{\left [ \, #1 \,; #2 \, \right ] }}
\newcommand{\intfo}[2]{\ensuremath{\left [ \, #1 \,; #2 \, \right [ }}

\newcommand{\intn}[2]{\ensuremath{[\![ \, #1 \,;\, #2 \,]\!]}}

\begin{document}

\begin{flushright}

\end{flushright}
\par \vskip .1in \noindent

\vspace{14pt}

\begin{center}
\begin{LARGE}
{\bf Long-distance asymptotic behaviour of multi-point correlation functions in massless quantum models}
\end{LARGE}

\vspace{30pt}

\begin{large}

{\bf N.~Kitanine}\footnote[1]{Universit\'{e} de Bourgogne, Institut de Math\'{e}matiques de Bourgogne, UMR 5584 du CNRS, France, Nicolai.Kitanine@u-bourgogne.fr},~~
{\bf K.~K.~Kozlowski}\footnote[2]{Universit\'{e} de Bourgogne, Institut de Math\'{e}matiques de Bourgogne, UMR 5584 du CNRS, France,
karol.kozlowski@u-bourgogne.fr},~~
{\bf J.~M.~Maillet}\footnote[3]{ Laboratoire de Physique, UMR 5672
du CNRS, ENS Lyon,  France,
 maillet@ens-lyon.fr},~~
{\bf V.~Terras}\footnote[4]{ Laboratoire de Physique, UMR 5672 du
CNRS, ENS Lyon,  France, veronique.terras@ens-lyon.fr}
\par

\end{large}

\vspace{40pt}

\centerline{\bf Abstract} \vspace{1cm}
\parbox{12cm}{\small  We provide a microscopic model setting that allows us to 
readily access to the large-distance asymptotic behaviour of multi-point correlation functions in  massless, one-dimensional, quantum models. The method of analysis we propose is based on the form factor expansion of the correlation functions and does not build on any field
theory reasonings. It constitutes an extension of the restricted sum techniques leading 
to the large-distance asymptotic behaviour of two-point correlation functions obtained previously. }
\end{center}

\vspace{40pt}

\section{Introduction}

Form factors expansions constitute  natural means for studying correlation functions in the case of massive integrable quantum field theories in the infinite volume  \cite{KarowskiWeiszFormFactorsFromSymetryAndSMatrices,SmirnovFormFactors}. Indeed, the presence of a gap in the spectrum leads both to regularity and well-ordering 
 of the form factor series as an effective asymptotic series in the large-distance regime. 
However, these two crucial properties are no-longer satisfied for massless models; 
their lack constituted an obstruction to an efficient formulation or even handling of form factor
series in the infinite volume limit of a model in the massless regime.
The main problem in the case of massless models is that the form factors of local operators vanish as a non-integer power-law    
in the model's volume $L$. This originates from  the operator's non-integer conformal dimension. 
When computing a form factor expansion, 
this non-integer power-law in $L$ vanishing of individual form factors should be compensated by certain multi-dimensional sums (which contain a number of summands diverging in $L$) over appropriate sub-classes  of low-lying excited states,
hence making this setting rather intricate to tackle.
 Recently, however, the authors, in collaboration with N. Slavnov,
have set forth a method of summing up, in the large-distance regime and for large but finite volume, the form factor expansion of two-point correlation functions
in massless quantum integrable models \cite{KozKitMailSlaTerRestrictedSums}. 
The matter is that, as we have observed, in the large-distance regime, the sums over form factors associated to 
scanning the whole spectrum of the theory localise to so-called restricted sums. The latter correspond, in physical terms, 
to summing up over all low-lying energy (of the order $1/L$, with $L$ being the volume assumed to be large but finite) excitations on the Fermi surface 
associated,  in the thermodynamic limit, to different Umklapp excitation momenta. 
Although the number of terms in 
these multiple restricted sums tends to infinity with the volume, it has been shown in \cite{KozKitMailSlaTerRestrictedSums} that they can be evaluated exactly using purely combinatorial  identities. Certain instances
of these identities already appeared in the context of harmonic analysis on the infinite 
dimensional symmetric group \cite{KerovOlshanskiVershikFirstOccurenceAbstracxtLevelMagicFormula} (see \cite{OlshanskiPointProcessAndInfiniteSymmetricGroup} for a first practical application to $z$-measures
on Young diagrams).

In this paper we continue developing the technique of summation, in the large-distance regime,
of form factor expansions, now for general $n$-point correlation functions. More precisely, we start from a model in \textit{large but finite} volume $L$
and make some technical assumptions on its spectrum and form factors. These properties are verified 
for quantum integrable models but we do trust that they should also hold for a much wider class of models. 
Building on these assumptions we develop summation techniques for form factor expansions that allow us to access to the large-distance regime of general multi-point correlation functions.  The corresponding restricted sums 
generalize those obtained for the two-point case \cite{KozKitMailSlaTerRestrictedSums}. In fact, in the case of general $n$-point correlation functions (with $n \ge 3$) the corresponding sums can be interpreted as multiple restricted sums of the previous type (obtained for two-point functions) however highly coupled between themselves hence rendering the needed summation identity quite non-intuitive. 
The derivation of this identity for the restricted sums corresponding to general $n$-point correlation functions  follows from the two possible representations
for the large-size asymptotic behaviour of a Toeplitz determinant generated by Fisher--Hartwig symbols:
the one corresponding to the Fisher--Hartwig asymptotic behaviour formula \cite{EhrhardtAsymptoticBehaviorOfFischerHartwigToeplitzGeneralCase} and the one
issuing from a form factor like expansion of this same Toeplitz determinant. 

The large-distance asymptotic summation of form factors method 
we develop does reproduce, within a microscopic approach to the 
model, the conformal field theory/Luttinger-Liquid predictions for the asymptotics of correlation functions in massless models. 
As an example, the 
large-distance behaviour of four point functions in the XXZ spin-1/2 chain is obtained as
\beq
\Big< \sg_{m_1}^{x} \cdot \sg_{m_2}^{x} \cdot \sg_{m_3}^{x} \cdot \sg_{m_4}^{x} \Big> \; \simeq \; 
2 \big|\mc{F}^{+}\big|^4  \bigg\{   
\Big| \f{  (m_2-m_1)\cdot (m_4-m_3)  }{  (m_3-m_1)   \cdot (m_4-m_1)\cdot (m_3-m_2) \cdot (m_4-m_2)   } \Big|^{\th} 
\; + \; (2 \leftrightarrow 3) \; + \; (2 \leftrightarrow 4)  \bigg\} \; +  \; \dots, 
\nonumber
\enq
matching the predictions given in \cite{GogolinNeversanTsvelikBozonizationTechniques,LutherPeschelCriticalExponentsXXZZeroFieldLuttLiquid}. Above, $\th$ is an explicit exponent that is expressed in terms of quantities parametrizing the low-lying
excitations of the XXZ spin-1/2 chain and the dots refer to higher order (sub-leading) terms. The amplitude $\mc{F}^{+}$ corresponds to a properly normalized 
in the length $L$ of the chain form factor of the $\sg^+$ operator taken between the model's overall ground state
$\ket{\Psi_g}$ and the local ground state in the sector with one spin being flipped down $\ket{\Psi_g^{(-1)}}$, 
\beq
\mc{F}^{+} \; = \; \lim_{L\tend + \infty} \bigg\{ \bigg(\f{L}{2\pi} \bigg)^{ \th } \bra{\Psi_g} \sg^+ \ket{\Psi_g^{(-1)}} \bigg\} \;. 
\nonumber
\enq
The critical exponent $\th$ was already present in the field theory predictions \cite{GogolinNeversanTsvelikBozonizationTechniques,LutherPeschelCriticalExponentsXXZZeroFieldLuttLiquid}. 
We do stress however that our method goes beyond in that it gives a direct access and interpretation of the amplitudes arising in front
of the algebraically decaying terms (along with the numerical pre-factor of 2).
It allows one readily to extract the large-distance asymptotic behaviour of any multipoint correlation function.
Moreover, our method is solely based on a genuine microscopic formulation of the model and at no stage of our analysis we do invoke a conjectural correspondence of the discrete 
model with some continuous field theory. 
Furthermore, with minor modifications, it should open a possibility to treat the case of 
time dependent correlation functions as it was done for two-point functions in \cite{KozKitMailSlaTerRestrictedSumsEdgeAndLongTime}.

We would like to stress that the general dependence in the distance we obtain is similar to the one 
for correlation functions of exponents of free fields, \text{cf} \cite{DostenkoFateevMultiPointFctsInCFT,DostenkoFateev4ptCorrFctCleq1CFT}. 
In fact, such a form already appeared in the late 60's paper of 
Kadanoff \cite{KadanoffFactPropsMultiPtsIsingCritical}. 
Recently, the same structure of the large-distance asymptotic behaviour has been proven, for the Ising correlation functions 
at the critical point, by Palmer \cite{PalmerRecoveryOfLuther-PeschelAsymptFromTauFctofSMJ} 
essentially based on the Sato, Miwa, Jimbo formalism \cite{JimMiwaSatoQuantumFieldsI,JimMiwaSatoQuantumFieldsIVGenMultiPtAnfFieldConstr,JimMiwaSatoQuantumFieldsVSomeExamples}. 
It is in fact in these pioneering works \cite{JimMiwaSatoQuantumFieldsI,JimMiwaSatoQuantumFieldsIVGenMultiPtAnfFieldConstr,JimMiwaSatoQuantumFieldsVSomeExamples} that the first rigorous approach to 
characterizing and computing multi-point correlation functions in free fermion model were given. 
The analysis developed there allowed these authors to provide, in particular,  a characterization of  
 multi-point correlation functions in a massless model 
for the case of the one-dimensional impenetrable Bose gas \cite{JimMiwaMoriSatoSineKernelPVForBoseGaz}
and to extract explicitly the large-distance asymptotic behaviour of the so-called one-particle reduced density
matrix in that model. 
Multi-point correlation functions in free fermion models
have been also investigated through Pfaffian or integrable integral operators methods. 
In \cite{AbrahamBarouchMcCoy4ptcomputationtoGetlongTimeforTwo}, Abraham, Barouch and McCoy 
derived a Pfaffian representation for four-point functions that allowed them to access to the 
time-dependent spin-spin correlation functions. 
In \cite{Bariev3ptFctsIsingOffCriticalAsympt,BarievmultiFctsIsingOffCriticalAsympt} Bariev derived the large-distance asymptotic behaviour 
of three and then multi-point correlation functions in the off-critical Ising model. 
In \cite{SlavnovPDE4MultiPtsFreeNLSM} Slavnov obtained a system of partial differential equation characterizing the 
$n$-point off-diagonal correlation functions in the one-dimensional impenetrable Bose gas.

The paper is organized as follows. In Section \ref{Section description des hypotheses sur le modele} we provide a definition of the microscopic model  and of all the quantities of interest to the problem. These definitions provide the general setting
that allows for an application of our form factor summation method. It surely does hold for quantum integrable models such as the XXZ spin-1/2 chain or the quantum non-linear Schr\"odinger model,
but should also be valid for some non-integrable models as well. Then, in Section \ref{Section calcul du DA generique}, we outline the main steps of our method and describe the overall features behind its philosophy. 
In particular we present the restricted sums of interest which are at the root of our approach to 
the large-distance asymptotic behaviour of the multi-point correlation function. Finally, we apply our general result to the case of certain integrable models in Section \ref{Section applications to integrable models}. 
Technical aspects related to the derivation of the multidimensional restricted sums relevant
for the present analysis are presented in Appendix \ref{Appendix summation identity}.

\section{Overall setting for the analysis of multi-point correlation functions}
\label{Section description des hypotheses sur le modele}

This paper aims at introducing a method of analysis of the large-distance asymptotic behaviour of $r$-point ground state correlation functions of the type 
\beq
C\big( \bs{x}_r; \bs{o}_r \big) \;  = \; \bra{ \Psi_g } \mc{O}_{1}(x_1)\dots \mc{O}_{r}(x_r) \ket{  \Psi_g } \;,
\label{definition multipoint corr fct}
\enq
where we agree to represent $r$-dimensional vectors as
\beq
\bs{x}_r \, = \, \big( x_1,\dots, x_r \big) \qquad \e{and} \qquad \bs{o}_r \, = \, \big( o_1,\dots, o_r \big) \;. 
\enq
On the level of \eqref{definition multipoint corr fct}, the correlation function $C\big( \bs{x}_r; \bs{o}_r \big)$ is defined for a model in finite volume $L$. In this formula, $\ket{  \Psi_g } $ represents the ground state of the model in finite volume 
whereas $\mc{O}_{a}(x)$ are some elementary (in the sense defined below) local operators
associated with the model and located at position $x$. Depending on the nature of the model (continuous or discrete), 
the position variables can be continuous or discrete. Throughout the paper, we 
shall build on the assumption that the states of the model are constructed out of 
a fixed number of pseudo-particles, this number possibly changing from one state to another. 
Within such a setting, we assume that one can associate a fixed  integer $o_a$ to each operator $\mc{O}_{a}(x)$. These integers refer to the jump in the pseudo-particle number of the states connected by the operator. More precisely, we assume that the operator 
$\mc{O}_{a}(x)$ is an elementary local operator that connects only states having $N + o_a$ and $N$ pseudo-particles, $N$ being arbitrary but $o_a$ fixed. In other words,
the only non-zero form factors of the operator $\mc{O}_{a}(x)$ are given by the matrix elements
\beq
\bra{ \Phi_N }  \mc{O}_{a}(x) \ket{  \Psi_{N+o_a} } 
\label{ecriture formule favteur de forme cadre general}
\enq
in which $ \ket{  \Psi_{N+o_a} } $ is some $N+o_a$ pseudo-particle eigenstate of the model's Hamiltonian whereas
$\ket{ \Phi_N }$ corresponds to an eigenstate built out of $N$ pseudo-particles. Any local operator can be decomposed in terms of a finite number of such elementary operators $\mc{O}_{a}(x)$. Finally, we focus solely 
on the case of translation invariant models (periodic boundary conditions) meaning that one can explicitly 
factor out the $x$-dependence out of \eqref{ecriture formule favteur de forme cadre general} in terms of the relative
momentum of the states $\ket{  \Psi_{N+o_a} } $ and $\ket{\Phi_N}$.

Our aim is to extract the large-distance asymptotic behaviour of the correlation function, in the infinite volume limit, 
$L \tend +\infty$, by using its form factor expansion obtained by inserting the closure relation in between the 
operators $\mc{O}_{a}(x)$. The matter is that the latter can be re-summed in the appropriate regimes of interest. 
In fact, we shall focus on the following regime of parameters 
\beq
1\;  \ll \;  |x_{l}-x_k| \cdot p_F \quad\e{where} \quad l,k=1, \dots, r-1, \ \ l \ne k, \qquad 
L \gg x_k  \quad k=1, \dots, r \;,
\enq
and $p_F$ denotes the Fermi momentum associated with the model. 

In the following, we describe a general framework such that, if the spectrum, eigenstates and form factors  
of a model fit within its grasp, then the present analysis holds. 
As we shall argue, this setting is verified for a number of quantum integrable models
such as the non-linear Schr\"{o}dinger model or the XXZ spin-1/2 chain in their massless phase. 
However, we are deeply convinced that the setting is, in fact, much more general and, in particular, 
encompasses also numerous instances of non-exactly solvable models.




\subsection{The spectrum}
\label{Subsection spectre model hypothesis}

We assume that, in the large-$L$ limit, the spectrum of the model is of particle-hole type
\footnote{Such an assumptions is not very restrictive as there are indications that the bound state part of the spectrum does not contribute 
to the algebraic part of the large-distance asymptotic behaviour of correlation functions. 
For instance, this is supported by exact calculations for the large-distance asymptotic behaviour of two-point functions 
in the XXZ spin-1/2 chain \cite{KozKitMailSlaTerXXZsgZsgZAsymptotics}. This is however most probably not the case for large-distance and large-time asymptotic behaviour.} and that the ground state $\ket{\Psi_g}$ is constructed within an $N$ pseudo-particle sector. We shall focus on states located in nearby sectors containing $N_s$ pseudo-particles, 
\beq
N_s \; = \; N \; + \; \ov{\bs{o}}_s \quad , \quad \ov{\bs{o}}_s \; = \; \sul{a=1}{s} o_a .
\label{definition Ns et os sum}
\enq
Within our setting, the states of the system in an $N_s$ pseudo-particle sector are labelled by two sets of $n$ integers associated to so-called particles and holes excitations, $n = 0, 1, \dots N_s$ :
\beq
p_1^{(s)} < \dots < p_n^{(s)} \qquad \e{and}  \qquad h_1^{(s)} < \dots < h_n^{(s)} \qquad \e{with} \qquad 
\left\{ \ba{c}  p_a^{(s)} \in  \intn{-M_{L}^{(1)} }{ M_{L}^{(2)} } \setminus \intn{1}{N_s}\, ,  \vspace{2mm} \\ 
				h_a^{(s)}  \in \intn{1}{N_s} \, .  \ea \right. 
\enq
The values taken by the integers $M_{L}^{(a)}$, $a=1,2$, strongly depend on the model. Typically, for models having no upper bound on their energy, one has $M_{L}^{(a)}=+\infty$, while
for model having an upper bound, $M_{L}^{(1)}, M_{L}^{(2)}$ are both finite but such that $M_{L}^{(a)}-N$, $a=1,2$,
both go to $+\infty$ sufficiently fast with $L$. Also note that throughout this paper
we shall adopt, for definiteness, the convention that the integers arising in the parametrization of a particle-hole excited 
state in the $N_s$ pseudo-particle sector always bear a superscript $(s)$.

A given choice of integers 
\beq
\mc{I}_{n}^{(s)} \; = \;\Big\{  \{ p_a^{(s)} \}_1^{n} \quad ;  \quad \{ h_a^{(s)} \}_1^{n} \Big\} 
\label{definition ensemble total pour parametriser etats}
\enq
for an excited state in the $N_s$ pseudo-particle sector gives rise to a set of rapidities $\{\wh{\mu}_{p_a}^{(s)} \}_1^n$ for the 
particle and $\{\wh{\mu}_{h_a}^{(s)} \}_1^n$ for the  holes associated with this state. The former and latter 
are computed from the integers $\mc{I}_{n}^{(s)}$ by the use of a so-called counting function
$\wh{\xi}_{ \mc{I}_{n}^{(s)} }$ as the unique solutions to 
\beq
\wh{\xi}_{ \mc{I}_{n}^{(s)} }\big( \wh{\mu}_{p_a}^{(s)} \big)  \; = \;  \f{ p_a^{(s)} }{L}  \qquad \e{and} \qquad 
\wh{\xi}_{ \mc{I}_{n}^{(s)} }\big( \wh{\mu}_{h_a}^{(s)} \big)  \; = \;  \f{ h_a^{(s)} }{L}\;. 
\label{ecriture equations definissant rap part et trous}
\enq
We do stress that the counting function depends on the choice of the excited state, \textit{ie} 
on the choice of integers $\mc{I}_{n}^{(s)}$. Hence, \eqref{ecriture equations definissant rap part et trous} is, in fact, a quite complicated set
of equations. We assume that all counting functions admit 
the large volume $L$ asymptotic expansions
\beq
\wh{\xi}_{ \mc{I}_{n}^{(s)} }\big(\om \big)  \; = \; \xi(\om) \; + \; \f{ 1 }{  L   }  \xi_{-1} (\om)
\; - \;  \f{ 1 }{  L   } F_{\mc{R}_{n}^{(s)}} (\om)   \; + \; \e{O} \Big( \f{ 1 }{ L^2 } \Big) \;. 
\label{ecriture definition fonction comptage}
\enq
The above asymptotic expansion involves three functions $\xi$, $\xi_{-1}$ and $F_{\mc{R}_{n}^{(s)}}$.  

\begin{itemize}
\item $\xi$ is the asymptotic counting function.  It is the same for \textit{all} excited states and defines a set of rapidities
$\{ \mu_a \}_{a \in \mathbb{Z} } $ as the unique solutions to the equation
\beq
\xi( \mu_a ) \; = \;  \f{ a }{ L }  \qquad \e{so} \;\; \e{that} \; \qquad 
\wh{\mu}_{p_a}^{(s)} \; \simeq \;  \mu_{ p_a^{(s)} }  \qquad \e{and} \qquad 
\wh{\mu}_{h_a}^{(s)} \; \simeq \;  \mu_{ h_a^{(s)} }  \;, 
\label{ecriture eqn Asymp pour position reseau rapidites}
\enq
at leading order in $L$.
\item  $F_{\mc{R}_{n}^{(s)} } $ is called the shift function (of the given excited state in respect to the model's ground state). 
It is a function of the macroscopic rapidities  
\beq
\mc{R}_{n}^{(s)} \; = \; \Big\{ \{ \mu_{ p_a^{(s)} } \}_1^n \; ; \; \{ \mu_{h_a^{(s)}} \}_1^n \Big\}
\enq
and also depends on the deviation $N_s - N = \ov{\bs{o}}_s$ of the number $N_s$ of pseudo-particles in the excited state
 in respect to $N$, the one for the ground state. 
\item $\xi_{-1}$ represents the $1/L$ corrections to the ground state's counting functions. 
It is this quantity that drives the non-trivial part of the first sub-leading corrections to the 
ground state's energy. It appears for normalisation purposes so that the shift function for the 
ground state vanishes, \textit{ie} $F_{\mc{R}_{0}^{(0)} } =0 $.  
\end{itemize}

In the large-$L$ limit and within such a setting, the rapidities for the ground state form a dense distribution 
 on $\intff{-q}{q}$ -- the so-called Fermi zone of the model -- with density 
$\xi^{\prime}$. The endpoints $\pm q$ are called the Fermi boundaries.
Further,  the observables of the model such as the relative momentum and energy 
are parametrized by the particle-hole rapidities
\beqa
\De\mc{E}\big( \mc{I}_{n}^{(s)}  \big) \; \equiv \; \mc{E}\big( \mc{I}_{n}^{(s)} \big) \; - \;  \mc{E}\big( \mc{I}_{0}^{(0)} \big) 
& = &  \sul{a=1}{n} \Big( \veps\big( \mu_{p_a^{(s)}} \big) \; - \;  \veps\big( \mu_{h_a^{(s)}} \big) \Big)
\; + \; \e{O}\Big( \f{1}{L} \Big)\, ,  \label{definition relative ex energy} \\
\De\mc{P}\big( \mc{I}_{n }^{(s)}  \big) \; \equiv \; \mc{P}\big( \mc{I}_{n}^{(s)} \big) \; - \;  \mc{P}\big( \mc{I}_{0}^{(0)} \big) & = & 
\sul{a=1}{n} \Big( p\big( \mu_{p_a^{(s)}} \big) \; - \;  p\big( \mu_{h_a^{(s)}} \big) \Big) 
\; + \; \e{O}\Big( \f{1}{L} \Big)\;.
\label{definition relative ex momentum}
\eeqa
Above, $\mc{I}_{0}^{(0)}=\{ \emptyset ; \emptyset \}$ refers to the set of integers which parametrizes the ground state
of the model and $\mc{P}\big( \mc{I}_{n}^{(s)} \big)$ and $\mc{E}\big( \mc{I}_{n}^{(s)} \big)$ are respectively the momentum and energy of the state parametrized by the set of integers $\mc{I}_{n}^{(s)}$.  Furthermore, $p$ is the so-called dressed momentum and similarly $\veps$ is the dressed energy. Typically, for quantum integrable models,
these functions are given as solutions to linear integral equations \cite{GaudinFonctionOndeBethe}. 
Their construction, for more complex, non-integrable models,
is \textit{a priori} much more complicated but carry direct physical meaning. Note that the particle-hole interpretation of the spectrum 
is manifest at the level of \eqref{definition relative ex energy}-\eqref{definition relative ex momentum}:
the excitations consist in adding "particles" with rapidities $\mu_{p_1^{(s)}}, \dots, \mu_{p_n^{(s)}}$
and associated energies $\veps\big(  \mu_{p_1^{(s)}} \big) , \dots, \veps\big( \mu_{p_n^{(s)}} \big)$
along with creating "holes" in the Fermi zone,  with rapidities $\mu_{h_1^{(s)}}, \dots, \mu_{h_n^{(s)}}$
and associated energies $-\veps\big(  \mu_{h_1^{(s)}} \big) , \dots, -\veps\big( \mu_{h_n^{(s)}} \big)$. 
The superscript $(s)$ refers to the fact that the excitation labelled by $\mc{I}_{n }^{(s)}$ takes place above the 
lowest lying energy level in the $N_s=N+\ov{\bs{o}}_s$ quasi-particle sector. 
We henceforth shall assume that the only roots on $\R$ of the dressed energy $\veps$ are at $\pm q$ and that the dressed momentum $p$
is a strictly monotonous function.

There is however a huge degeneracy in what concerns the $\e{O}(1)$ part of the excitation energy 
$\De\mc{E}\big( \mc{I}_{n}^{(s)}  \big) $: numerous choices of the set $\mc{I}_n^{(s)}$
will lead to the \textit{same} $\e{O}(1)$ part of  $\De\mc{E}\big( \mc{I}_{n}^{(s)}  \big) $. For instance, it follows
from \eqref{ecriture definition fonction comptage}-\eqref{ecriture eqn Asymp pour position reseau rapidites} that, 
given any integer $\ell$, 
\beq
\mu_{k} - \mu_{k+\ell} \; = \; \e{O}\big( \f{\ell}{L}\big)\, , \qquad \e{so} \; \e{that} \qquad 
\mc{I}_n^{(s)} \; = \; \Big\{ \{p_a\}_1^n \; ; \; \{h_a\}_1^n \Big\} \quad \e{and} \quad 
\wt{\mc{I}}_n^{(s)} \; = \; \Big\{ \{p_a + k_a \}_1^n \; ; \;  \{h_a+t_a\}_1^n \Big\}
\enq
where $k_a, t_a$ are some $L$-independent integers, will lead to the same value for the $\e{O}(1)$
part of $\De\mc{E}$, \textit{ie}
\beq
\De\mc{E}\big( \mc{I}_{n}^{(s)}  \big) \; - \;  \De\mc{E}\big( \wt{\mc{I}}_{n}^{(s)}  \big)  \; = \; \e{O}\Big( \f{1}{L} \Big) \;. 
\enq
%
%
%




\subsection{The form factors}

In the analysis of the multipoint correlation function $C\big( \bs{x}_r; \bs{o}_r \big)$
\eqref{definition multipoint corr fct} through its form factor expansion, one needs to 
access to the form factors of the local operators $\mc{O}_s$, $s=1,\dots,r$,  between 
two excited states of the model. According to our assumptions on the structure of the model's spectrum, 
these
will be labelled by two sets of multi-indices $\mc{I}_{ m }^{(s-1)} $ and $\mc{I}_{ n }^{(s)}$
corresponding to the outgoing (bra) and ingoing (ket) states. In fact, since we imposed periodic boundary conditions
on the model and  required translation invariance, the form factors will satisfy 
\beq
\bra{ \Psi\big( \mc{I}_{ m }^{(s-1)} \big) } \mc{O}_s(x)   \ket{ \Psi\big( \mc{I}_{ n }^{(s)} \big) } \; = \; 
\ex{ i x (\De\mc{P})_{s-1}^{s}  }
 \bra{ \Psi\big( \mc{I}_{ m }^{(s-1)} \big) } \mc{O}_s(0)   \ket{ \Psi\big( \mc{I}_{ n }^{(s)} \big) }, \ \ 
 (\De\mc{P})_{s-1}^{s}  \; = \; \mc{P}\big({\mc{I}_{m }^{(s-1)} }\big) - \mc{P}\big({\mc{I}_{ n }^{(s)} }\big)  \;. 
\enq
Scalar observables introduced in the last subsection were parametrized, in the large-$L$ limit, 
solely by the macroscopic set of rapidities $\mc{R}_{n}^{(s)}$ subordinate to the set of multi-indices $\mc{I}_{n}^{(s)}$. 
This is no longer the case for form factors as we demonstrated in our previous work  \cite{KozKitMailSlaTerThermoLimPartHoleFormFactorsForXXZ}. Within our setting, the latter are parametrized, in the large-$L$ limit, 
\textit{not only by} the sets of macroscopic rapidities $\mc{R}_{ n }^{(s)}, \mc{R}_{ m }^{(s-1)} $ ,
but also by the sets of \textit{discrete} integers $\mc{I}_{n}^{(s)}, \mc{I}_{ m }^{ (s-1) }$. 
Namely, for properly normalized states $\ket{ \Psi\big( \mc{I}_{ n }^{(s)} \big) } $ and 
their duals $\bra{ \Psi\big( \mc{I}_{ m }^{(s-1)} \big) } $, the form factors take the form 
\bem
\mc{F}_{\mc{O}_s}\big( \mc{I}_{ m }^{(s-1)} \mid \mc{I}_{ n }^{(s)}  \big) \; = \;
\bra{ \Psi\big( \mc{I}_{m }^{(s-1)} \big) } \mc{O}_s(0)   \ket{ \Psi\big( \mc{I}_{ n }^{(s)} \big) }  \\
\; = \; 
\mc{S}^{( \mc{O}_s)} \Big(   \mc{R}_{ m }^{(s-1)} \, ; \,   \mc{R}_{n}^{(s)}     \Big)
     \cdot 	\mc{D}^{(s)}
 \Big(   \mc{R}_{m}^{(s-1)}\, ; \,  \mc{R}_{n}^{(s)}     \big| \, 
    \mc{I}_{ m }^{(s-1)} \, ; \,   \mc{I}_{ n }^{(s)}   	\Big)	  \cdot 
\bigg(  1+ \e{O}\Big( \f{\ln L }{ L } \Big) \bigg)      		 \;. 
\end{multline}
In this decomposition 
\begin{itemize}
\item $\mc{S}^{( \mc{O}_s )}$ is called the smooth part since it only depends 
on the macroscopic rapidities $\mc{R}_{n}^{(s)}, \mc{R}_{ m }^{(s-1)} $, and this in a smooth manner. 
As a consequence, a small (of the order $\e{O}(1)$) change in the value of the integers
parametrizing the state, say $p_a \hookrightarrow p_a + \kappa$, will not change the value of the 
smooth part (up to 1/$L$ corrections).  Furthermore, being a set function, it is invariant under permutation
of the particle or hole rapidities. 
\item The smooth part enjoys hypergeometric-like reduction properties. Namely, if within a given set of macroscopic rapidities   
a particle's rapidity coincides with a hole rapidity, then this dependence simply disappears. More precisely
\beq
\mc{S}^{( \mc{O}_s)} \Big(  \mc{R}_{ m }^{(s-1)} ; \,  \mc{R}_{ n }^{(s)}    \Big)
_{\mid \mu_{p_k^{(s)}}= \mu_{h_{\ell}^{(s)}} } \; = \; 
\mc{S}^{( \mc{O}_s)} \Big(  \mc{R}_{ m }^{(s-1)} ; \,  \wh{\mc{R}}_{ n }^{(s)}    \Big)
\qquad \e{with} \qquad 
\wh{\mc{R}}_{n}^{(s)}  \; = \; \Big\{ \{ \mu_{ p_a^{(s)} } \}_{1; \not=k}^{ n } 
\; ; \; \{ \mu_{h_a^{(s)}} \}_{1; \not= \ell}^{ n } \Big\} \;. 
\nonumber
\enq
The same type of reduction holds as well for particle-hole rapidities associated with the excited states in the 
$N_{s-1}$ quasi-particle sector. 

\item $\mc{D}^{(s)}$ is called the discrete part since it depends not only on the 
macroscopic rapidities $\mc{R}_{ n }^{(s)}, \mc{R}_{ m }^{(s-1)} $ but also 
\textit{explicitly} (\textit{ie} not through the parametrization by the macroscopic rapidities)
on the sets of integers labelling the excited states
$\mc{I}_{ n }^{(s)}, \mc{I}_{ m }^{(s-1)} $. 
The main effect of such a dependence is that a small change in the value of the
integers parametrizing the state does imply a significant change (of the order of $\e{O}(1)$) in the
value of $\mc{D}^{(s)}$. The discrete part thus keeps track of the microscopic details of the different
excited states. 
\end{itemize}
The smooth part represents, in fact, a non-universal part of the model's form factors. 
Its explicit expressions not only depends on the operator $\mc{O}_s$ but also varies strongly 
from one model to another, see \cite{KozKitMailSlaTerThermoLimPartHoleFormFactorsForXXZ,KozFFConjFieldNLSELatticeSpacingGoes0} 
for examples issuing from quantum integrable models. However, the part $\mc{D}^{(s)}$
is entirely universal within the present setting of the description of the model's spectrum. 
It solely depends on the values of the pseudo-particle sectors that the operator connects, \textit{viz}. the integer $o_s=N_s-N_{s-1}$. 
Its general explicit expression plays no role in our analysis, in the sense that we shall only 
need the expression for specific excited states, namely the one belonging to the so-called $\ell_s$-critical classes that we define below. 

We refer the interested reader to \cite{KozKitMailSlaTerThermoLimPartHoleFormFactorsForXXZ,KozKitMailSlaTerRestrictedSums} for a thorougher discussion relative to the 
origin of the discrete $\mc{D}^{(s)}$ and smooth $\mc{S}^{(\mc{O}_s)}$ parts in the framework of Bethe Ansatz 
solvable models.

%
%

\subsection{The critical $\ell_s$ class}

As we shall argue in the following, only a very specific class of excited states
will play an effective role in our analysis -- the so-called critical states. 
These excited states are characterized by the fact that, in the $L \tend +\infty$ limit, 
\textit{all} macroscopic rapidities describing the particle and hole excitations "collapse" on the model's Fermi boundary:
\beq
\mu_{p_a^{(s)}} \simeq   \pm q \qquad \e{and} \qquad \mu_{h_a^{(s)}} \simeq   \pm q  \;. 
\enq
There, the $\pm$ sign depends on whether the particle or hole's rapidity collapses on the right or left Fermi boundary.

A set of integers $\mc{I}^{(s)}_n$ is said to parametrize 
a critical excited state if the associated particle-hole integers $\{ p_a^{(s)} \}_1^{ n }$ and $\{ h_a^{(s)} \}_1^{ n }$  can be represented as
\beq
\big\{ p_a^{(s)}  \big\}_1^{ n } \; = \;  \big\{  N_s + p_{a;+} \big\}_1^{ n_{p;+} } \, \cup  \,
\big\{ 1 -  p_{a;-} \big\}_1^{n_{p;-}}  \qquad \e{and} \qquad 
 \big\{ h_a^{(s)} \big\}_1^{n } \; = \;  \big\{1+ N_s - h_{a;+} \big\}_1^{n_{h;+}} \, \cup  \, 
\big\{ h_{a;-}  \big\}_1^{ n_{h;-} }   ,
\label{ecriture decomposition locale part-trou close Fermi zone}
\enq
where the integers $p_{a;\pm}, h_{a;\pm} \in \mathbb{N}^{*}$ are "small" compared to $L$, \textit{ie}
\beq
\lim_{L\tend +\infty} \f{ p_{a;\pm} }{ L }  \; = \; \lim_{L\tend +\infty} \f{ h_{a;\pm} }{ L }  \; = \; 0 \;, 
\enq
 and the integers $n_{p/h;\pm}$ satisfy to the constraint
\beq
n_{p;+} \, + \,  n_{p;-} \; = \;  n_{h;+}\, + \,  n_{h;-} \;= \; n \;. 
\enq
Within this setting, one can readily check that the critical excited state described above 
will have $n_{p;+/-}$ particles, resp. $n_{h;+/-}$ holes, on the right/left end of the Fermi zone $\intff{-q}{q}$
associated with the $N_s$ pseudo-particle sector. 

In fact, one can distinguish between various critical states by organizing them into so-called 
$\ell_s$-critical classes. This classification takes its origin in the fact that all such states have a vanishing 
excitation energy (up to \e{O}(1/$L$) corrections) but can be gathered into classes depending on the value of their
macroscopic momenta $2\ell_s p_F$, where $p_F=p(q)$ is the so-called Fermi momentum and 
\beq
\ell_s \; = \;  n_{p;+}  \; - \;  n_{h;+} \; = \; n_{h;-}  \; - \;  n_{p;-} \;. 
\label{ecriture lien shift ells et differences part trous sur bords zone Fermi}
\enq
The index $s$ in $\ell_s$ is there so as to keep track of the $N_s$ pseudo-particle sector with which the
critical class is associated.  However, for the sake of simplifying the notations in formulae, we chose not to emphasize the $s$ dependence
in the left or right particle/hole numbers $n_{p/h;\pm}$. 

In fact, one has the following large-$L$ expansion for the relative excitation momentum \eqref{definition relative ex momentum}  
associated with the $\ell_s$-critical class excited state described above
\beq
\De\mc{P}\big( \mc{I}_{n }^{(s)}  \big) \; =\;  2 \ell_s  p_F 
\; + \; \f{2\pi}{L} \Bigg\{ \sul{a=1}{ n_{p;+} } p_{a;+} \; + \;  \sul{ a=1 }{ n_{h;+} } (h_{a;+}  -1 ) \Bigg\}
\; - \; \f{2\pi}{L} \Bigg\{ \sul{a=1}{n_{p;-} } (p_{a;-} -1)\; + \;  \sul{a=1}{n_{h;-}}  h_{a;-}  \Bigg\}  \; + \; ...
\label{ecriture ex momentum pour etats ell shiftees}
\enq
where we \textit{do stress} that all the terms included in the dots are either of the order of $\e{O}(1/L)$  \textit{but} 
do \textit{not} depend on the integers $p_{a;\pm}$ and $h_{a;\pm}$ or they depend 
on these integers \textit{but} are of the order of $\e{O}(1/L^2)$. 

It is convenient, for such $\ell_s$-critical excited states, to use a reparametrisation of the sets of integers labelling the excitations. 
It is readily seen that the excited states of an $\ell_s$-critical class are described by two sets of integers
$\mc{J}^{(s)}_{ n_{p;\pm}; n_{h;\pm} }$, each containing all the informations on the local 
integers on the right ($+$) or left ($-$) Fermi boundaries. Here, we agree upon
\beq
  \mc{J}^{(s)}_{  n ; m  }  \; = \; \Big\{ \{ p_{a}^{(s)} \}_1^{ n } 
\; ; \;  \{ h_{a}^{(s)} \}_1^{ m } \Big\}  \;. 
\enq
Note that the value of $\ell_s$ is encoded in the very notation $\mc{J}^{(s)}_{ n_{p;\pm}; n_{h;\pm} }$ 
by means the identification \eqref{ecriture lien shift ells et differences part trous sur bords zone Fermi}. 
Since there is a one-to-one correspondence between the set $\mc{I}^{(s)}_{n} $ and 
$\mc{J}^{(s)}_{ n_{p;+}; n_{h;+} } \cup \mc{J}^{(s)}_{ n_{p;-}; n_{h;-} }$,
we shall identify the two sets. We do stress that the parameter $s$ does play a role
in this correspondence, \textit{cf} \eqref{ecriture decomposition locale part-trou close Fermi zone}.

\subsection{Large-$L$ expansion of form factors connecting critical states}

As we have already mentioned, solely critical states enter in the analysis of the 
form factor expansion at large spacial separation between the operators. Thus, 
we now present the explicit expression for the form factor taken between such states. 
Given two excited states
\beq
\mc{I}^{(s-1)}_m \; \equiv \; \mc{J}^{(s-1)}_{ m_{p;+}; m_{h;+} } \cup \mc{J}^{(s-1)}_{ m_{p;-}; m_{h;-} } 
\quad \e{and} \quad 
\mc{I}^{(s)}_n \; \equiv \; \mc{J}^{(s)}_{ n_{p;+}; n_{h;+} } \cup \mc{J}^{(s)}_{ n_{p;-}; n_{h;-} } 
\enq
 belonging respectively to the 
\beq
 \ell_{s-1}\; = \; m_{p;+} -  m_{h;+} \; = \; m_{h;-} -  m_{p;-}   \quad \e{and} \quad
 \ell_{s}\; = \; n_{p;+} -  n_{h;+} \; = \; n_{h;-} -  n_{p;-}
\enq
 critical classes, we shall assume that the form factors of local operators take the form\footnote{A priori, 
this formula could be modified by the presence of a sign factor $(-1)^{\sg_s+\sg_{s-1}}$ originating from the determination of the 
square root one chooses for the norms. The integer $\sg_{s}$ depends, \textit{a priori}, on $\mc{I}_n^{(s)}$. However, the very structure 
of such a sign factor makes its contribution to the form factor expansion irrelevant as it always appears twice. 
We have therefore disregarded its presence from the very beginning.
One can also allow for the multiplication of \eqref{ecriture conjecture form general facteur de forme deux etats excites} by a $L$-dependent phase factor of the form $e^{i\pi L(\phi_{s}-\phi_{s-1})}$, with $\phi_s$ depending on  $\mc{I}_n^{(s)}$.  Once again, the appearance of such a phase factor has no influence on the form factor expansion \eqref{ecriture asympt dominantes comme serie part trous}, hence we simply omit it here (note however that the presence of such a phase factor should be taken into account in the definition \eqref{ecriture definition facteur de forme proprement normalise macroscopique} of $\mc{F}_{\mc{O}_s}(\ell_{s-1}, \ell_s)$). } 
\bem
\mc{F}_{\mc{O}_s}\bigg(  \mc{I}^{(s-1)}_m  \Big| \mc{I}^{(s)}_n \bigg)  \; = \; 
\mc{F}_{\mc{O}_s}(\ell_{s-1}, \ell_s)  \cdot C^{(\ell_{s-1}; \ell_s)} \big( \nu_s^+, \nu_s^- \big) \\
\times 
\msc{F}^{(+)}\Big[  \mc{J}^{(s-1)}_{ m_{p;+}; m_{h;+}  } ;  \mc{J}^{(s)}_{ n_{p;+} ; n_{h;+} }     \mid \nu_s^{+} \Big] \cdot 
 \msc{F}^{(-)}\Big[ \mc{J}^{(s-1)}_{ m_{p;-} ; m_{h;-} } ;  \mc{J}^{(s)}_{ n_{p;-} ; n_{h;-} }     \mid 
\nu_s^{-} \Big] \;. 
\label{ecriture conjecture form general facteur de forme deux etats excites}
\end{multline}

The constituents of the above formula are parametrized by the values 
\beq
\nu_{s}^{+} \; =  \; \nu_s(  q )  \, - \, o_s \qquad \e{and} \qquad
\nu_{s}^{-} \; =  \; \nu_s(  -q )
\enq
that the relative shift function between the $\ell_{s}, \ell_{s-1}$ critical states,
\beq
\nu_{s}(\la) \; = \;  F_{s-1}(\la) \; - \; F_{s}(\la) \, ,
\enq
takes on the right/left endpoints of the Fermi zone, up to subtracting the level $o_s$ of the operator $\mc{O}_s$
in the case of the right endpoint. 
In formula \eqref{ecriture conjecture form general facteur de forme deux etats excites}, 
all the pre-factors but $\msc{F}_{\mc{O}_s}(\ell_{s-1}, \ell_s) $ issue from the "relevant" part of the discrete term. 
Their explicit expressions are given below. Even though these may appear as slightly complicated, the physical
interpretation of each factor in \eqref{ecriture conjecture form general facteur de forme deux etats excites} is crystal clear. 
Indeed, the quantity $\msc{F}_{\mc{O}_s}(\ell_{s-1}, \ell_s)$ 
represents the properly normalized finite and non-universal (\textit{ie} model and operator dependent) 
part of the large-$L$ behaviour of the 
form factor of the operator $\mc{O}_s$ taken between fundamental representatives of the $\ell_s$ and $\ell_{s-1}$ critical classes. 
More precisely, it is defined as
\beq
\mc{F}_{\mc{O}_s}(\ell_{s-1}, \ell_s) \; = \; \lim_{L\tend +\infty}\Bigg\{
 \bigg( \f{ L }{ 2\pi } \bigg)^{ \rho_s(\nu_s^{+}) + \rho_s(\nu_s^-) }
\bra{ \Psi\big( \mc{L}_{ \ell_{s-1} }^{(s-1)} \big) } \mc{O}_s(0)   \ket{ \Psi\big( \mc{L}_{ \ell_{s} }^{(s)} \big) } \Bigg\}
\label{ecriture definition facteur de forme proprement normalise macroscopique}
\enq
in which the sets of integers $\mc{L}_{ \ell_{s-1} }^{(s-1)}$  and $ \mc{L}_{ \ell_{s} }^{(s)} $ 
parametrizing the excited states correspond to the fundamental representatives of the $\ell_{s-1}$ and $\ell_{s}$ critical classes. 
Namely,   $ \mc{L}_{ \ell_{s} }^{(s)} $  is the set of particle-hole integers living on the Fermi boundary 
in the $N_s$ pseudo-particle sector
such that 
\beq
 \mc{L}_{ \ell_{s} }^{(s)}  \; = \; 
\left\{   \ba{cc} \Big\{  \{ p_{a;+}  = a \}_1^{\ell_s} \; ;  \; \{ \emptyset \}  \Big\} \bigcup
\Big\{  \{ \emptyset \} \; ; \;  \{ h_{a;-} = a \}_1^{\ell_s}  \Big\}  &
				\e{if} \; \ell_s \geq 0   \vspace{2mm} \\ 
		\Big\{  \{ \emptyset \} \; ; \;  \{ h_{a;+}  = a \}_1^{-\ell_s}   \Big\} \bigcup
\Big\{    \{ p_{a;-} = a \}_1^{-\ell_s}  \; ; \; \{ \emptyset \}  \Big\}   &
				\e{if} \; \ell_s \leq 0 \ea \right. 		\; . 
\enq
The power of the volume $L$ arising in \eqref{ecriture definition facteur de forme proprement normalise macroscopique} involves
the right $\rho_{s}(\nu_s^+)$ and left $\rho_{s}(\nu_s^-)$ scaling dimensions whose generic expression reads  
\beq
 \rho_{s}(\nu) \; = \;  \f{1}{2}  (\ell_s-\ell_{s-1})^2 \; + \; \f{1}{2} \nu^2 
\; - \; (\ell_s-\ell_{s-1})\nu \;. 
\enq

The factor $\msc{F}^{(+)}$ corresponds to the contributions of the excitations 
on the right Fermi boundary of the model. It is a function of the value taken on the 
right Fermi boundary by the relative shift function $\nu_s$ associated with the $\ell_s$ class of interest. 
Also, $\msc{F}^{(+)}$ depends on the sets of integers $\mc{J}^{(s-1)}_{ m_{p;+}; m_{h;+} } $ and  
$\mc{J}^{(s)}_{ n_{p;+}; n_{h;+} } $ parametrizing the excitations on the right boundary for the $s-1$ and $s$ excited states. 
Given two sets of integers  
\beq
\mc{J}_{ n_{p}; n_{h} } \; = \; \Big\{  \{p_a\}_1^{n_p}  \; ; \; \{h_a\}_1^{n_h} \Big\} \qquad \e{and} \qquad 
\mc{J}_{ n_{k}; n_{t} } \; = \; \Big\{  \{k_a\}_1^{n_k}  \; ; \; \{t_a\}_1^{n_t} \Big\}  \;,
\label{definition ensembles part trous modeles}
\enq
the right Fermi boundary critical form factor reads 
\bem
\msc{F}^{(+)}\Big[  \mc{J}_{ n_{p}; n_{h} } ;  \mc{J}_{ n_{k}; n_{t} }     \mid \nu \Big]
 \; = \; \bigg( \f{ 2\pi }{ L } \bigg)^{ \rho_{s}(\nu) }  \hspace{-2mm}
%
%
\cdot (-1)^{n_{t} } \cdot \bigg( \f{\sin [\pi \nu] }{ \pi } \bigg)^{ n_{t} \, + \, n_{h}  }  
\cdot \varpi\Big( \mc{J}_{ n_{p}; n_{h} } ;  \mc{J}_{ n_{k}; n_{t} }    \mid \nu \Big)  \\
\times \f{ \pl{a>b}{ n_{p} } (p_{a} - p_{b} ) \cdot \pl{a<b}{ n_{h}  } ( h_{a} - h_{b} ) }
{ \pl{a=1}{ n_{p} } \pl{b=1}{ n_{h} } (p_{a} + h_{b} - 1) }  \cdot 
\f{ \pl{a>b}{ n_{k} } (k_{a} - k_{b} ) \cdot \pl{a<b}{ n_{t}  } ( t_{a} - t_{b} ) }
{ \pl{a=1}{ n_{k} } \pl{b=1}{ n_{t} } (k_{a} + t_{b} - 1) }   \cdot 
\Ga\Bigg( \ba{cccc}  
\big\{ p_{a} + \nu \big\}_1^{n_p} , & \big\{h_{a}-\nu \big\}_1^{n_h} 
, & \big\{ k_{a}-\nu   \big\}_1^{n_k}  , & \big\{ t_{a} +\nu  \big\}_1^{n_t}  \\
 \big\{ p_{a} \big\}_1^{n_p}  , & \big\{ h_{a} \big\}_1^{n_h} , & \big\{ k_{a} \big\}_1^{n_k} , &  \big\{ t_{a} \big \}_1^{n_t} 
    \ea \Bigg) \;. 
\nonumber
\end{multline}
Note that $\mc{F}^{(+)}$ decays as an algebraic power of the volume $L$ with a 
scaling dimension $ \rho_{s}(\nu)$ . 
Likewise, the factor $\msc{F}^{(-)}$ corresponds to the contributions of the excitations 
on the left Fermi boundary of the model. It is a function of the value $\nu_s^{-}$ taken on the 
left Fermi boundary by the relative shift function $\nu_s$ and of the sets
$\mc{J}^{(s-1)}_{ m_{p;-}; m_{h;-} } $ and  $\mc{J}^{(s)}_{ n_{p;-}; n_{h;-} } $ of integers parametrizing the 
excitations on the left Fermi boundaries for the $s-1$ and $s$-excited states. 
Given sets as in \eqref{definition ensembles part trous modeles} for the parametrization of 
the excitations on the left Fermi boundary, one has 
\bem
\msc{F}^{(-)}\Big[  \mc{J}_{ n_{p}; n_{h} } ;  \mc{J}_{ n_{k}; n_{t} }     \mid \nu \Big] 
\; = \; \bigg( \f{ 2\pi }{ L } \bigg)^{\rho_{s}(\nu) } \hspace{-3mm}
%
\cdot (-1)^{n_{h} } \cdot \bigg( \f{\sin [\pi \nu] }{ \pi } \bigg)^{ n_{h} \, + \, n_{t}  }  
\cdot \varpi\Big( \mc{J}_{ n_{p}; n_{h} } ;  \mc{J}_{ n_{k}; n_{t} }     \mid - \nu \Big) \\
\hspace{-1cm} \times \f{ \pl{a>b}{ n_{p} } (p_{a} - p_{b} ) \cdot \pl{a<b}{ n_{h}  } ( h_{a} - h_{b} ) }
{ \pl{a=1}{ n_{p} } \pl{b=1}{ n_{h} } (p_{a} + h_{b} - 1) }  \cdot 
\f{ \pl{a>b}{ n_{k} } (k_{a} - k_{b} ) \cdot \pl{a<b}{ n_{t}  } ( t_{a} - t_{b} ) }
{ \pl{a=1}{ n_{k} } \pl{b=1}{ n_{t} } (k_{a} + t_{b} - 1) } 
 \cdot 
\Ga\Bigg( \ba{cccc}  
 \big\{p_{a}-\nu \big\} , & \big\{ h_{a}+\nu \big\}  , & \big\{ k_{a} + \nu \big\} 
 , & \big\{  t_{a} - \nu \big\} 
   \\
\big\{ p_{a} \big\} , & \big\{ h_{a} \big\} , & \big\{ k_{a} \big\}  , & \big\{ t_{a} \big\}\ea \Bigg)
\;.   
\nonumber
\end{multline}
Just as for the right Fermi boundary critical form factor, the left one decays algebraically
with the volume of the model with a left scaling dimension $\rho_{s}(\nu)$.

The expressions for the right and left Fermi boundary factors involve products of $\Ga$-functions written as
\beq
\Ga \Bigg(  \ba{c} \{v_a\}_1^n \\ \{w_a\}_1^m \ea \bigg) \; = \; 
\Ga \Bigg(  \ba{c} v_1, \dots , v_n  \\ w_1, \dots , w_m \ea \bigg) \; = \;  \f{\pl{a=1}{n} \Ga(v_a) }{ \pl{a=1}{m} \Ga(w_a) }  \;. 
\label{introduction notation produit compact gamma functions}
\enq
The $\Ga$ functions encode the structure of excitations within a given $(s)$-sector of excitations on the Fermi zone. 
The expressions for the left and right local factors $\msc{F}^{(\pm)}$ also involve the function $\varpi$ whose presence translates 
the interactions between 
integers parametrizing the neighbouring sectors $(s-1)$ and $(s)$ that are connected by the operator $\mc{O}_s$. 
Taken two sets of integers as in \eqref{definition ensembles part trous modeles}, one has 
\beq
 \varpi\Big( \mc{J}_{ n_{p}; n_{h} } ;  \mc{J}_{ n_{k}; n_{t} }    \mid \nu \Big)  
			\; = \;  \pl{a=1}{n_h  }   \Bigg\{  \f{  \pl{b=1}{ n_k } \big( 1-k_b-h_a+\nu  \big)    }
{  \pl{b=1}{n_t } \big( t_b-h_a+\nu \big)  }  \Bigg\}  \cdot 
\pl{a=1}{n_p}  
\Bigg\{  \f{ \pl{b=1}{n_t} \big(p_a+t_b+\nu  -1 \big)    }
{  \pl{b=1}{n_k} \big( p_a - k_b + \nu \big)  }  \Bigg\}		\;. 
\label{definition de la fonction  varpi}
\enq

 Finally, the function $C^{(\ell_{s-1}; \ell_s)} \big( \nu_s^+, \nu_s^- \big)$ is a normalization constant. 
It provides the appropriate normalization of the expression for the general critical form factor. 
It is such that it precisely cancels out the $L$-independent contributions of the right and left critical form factors
when focusing on the fundamental representative of the $\ell_s$, $\ell_{s-1}$ critical classes. 
Its explicit form can be computed in terms of the Barnes $G$-function \cite{BarnesDoubleGaFctn1} and reads:
\beq
C^{(\ell_{s-1}; \ell_s)} \big( \nu_s^+, \nu_s^- \big) \; = \; 
  G\bigg( \ba{cc} 1 + \nu_s^- , 1-\nu_s^+ , 1+\ell_{s} - \nu_{s}^-, 1-\ell_s + \nu_s^+  \\
1 -  \ell_s + \nu_s^- ,   1+\ell_s-\nu_s^+   , 1 - \ell_{s-1}+\ell_s - \nu_s^- , 1+ \ell_{s-1}-\ell_s + \nu_s^+  	\ea \bigg)   
	\;. 	
\enq
Note that above, we have adopted similar product conventions as for $\Ga$-functions, \textit{cf} 
\eqref{introduction notation produit compact gamma functions}.




\section{Asymptotic behaviour of multi-point correlation functions}
\label{Section calcul du DA generique}

\subsection{Long-distance asymptotic expansion of multi-point functions}

In this section, building on our previous assumptions on the structure of the spectrum and form factors
of local operators, we derive the large-distance asymptotic behaviour of the correlator $C\big( \bs{x}_r; \bs{o}_r \big)$, \textit{cf}.
\eqref{definition multipoint corr fct}. 
As already discussed, our method builds on a microscopic analysis of the form factor expansion for this
correlator. Hence, in between the operators $s$ and $s+1$ we insert a decomposition of the identity. 
In principle, we should thus sum up over all the states of the model, namely
the states generated by the particle-hole excitations along with the more complex states related to bound states. 
However, building on the assumption that states containing 
bound states solely generate exponentially small corrections to the large-distance regime in the $|x_{k+1}-x_k|\cdot p_F \gg 1$ limit, 
we only sum-up over the particle-hole states, as described in Section \ref{Subsection spectre model hypothesis}. 
Thus, for each $s$ ($s=1,\dots,r-1$) we sum up over all possible choices of the sets 
$\big\{ \mc{I}_{ n^{(s)} }^{(s)} \big\}$ as given in \eqref{definition ensemble total pour parametriser etats},
with $n^{(s)}=0,1,\dots, $. In other words the multi-point correlator 
is recast as 
\beq
C\big( \bs{x}_r; \bs{o}_r \big) \;  = \; \pl{s=1}{r-1} \bigg\{ \sul{ \big\{ \mc{I}_{n^{(s)}}^{(s)} \big\}  }{}  \bigg\}
\; \cdot \; \pl{s=1}{r-1} \bigg\{  \exp\Big[ i(x_{s+1}-x_s) \cdot \De \mc{P} (\mc{I}_{n^{(s)}}^{(s)}) \Big] \bigg\} \cdot 
\pl{s=1}{r} \mc{F}_{\mc{O}_s}\Big( \mc{I}_{n^{(s-1)}}^{(s-1)}  \mid  \mc{I}_{n^{(s)}}^{(s)}    \Big) \;,
\label{ecriture developpement multipoint sur etats particules trous}
\enq
in which $\De \mc{P}  \big(\mc{I}_{n^{(s)}}^{(s)}\big)$ stands for the relative excitation momentum defined
in \eqref{definition relative ex momentum}. 

We do stress that the equality \eqref{ecriture developpement multipoint sur etats particules trous} is
to be understood in the sense of having neglected all the contributions from the bound-states to the 
form factor expansion. 

We now follow exactly the reasoning outlined in \cite{KozKitMailSlaTerRestrictedSums}. The
multidimensional sum in \eqref{ecriture developpement multipoint sur etats particules trous} contains
quickly oscillating terms with speed measured by the magnitude of  $|x_{k+1}-x_k|\cdot p_F$, $k=1,\dots, r-1$. 
Thus, by analogy with a multiple integral of oscillatory type, the leading asymptotic behaviour 
will issue from localising the sums at the 
endpoints of the summation domain or at the saddle-points of the oscillating exponent. 
However, due to monotonicity of the dressed momentum function $p$, the relative excitation momentum
$\De \mc{P}  \big(\mc{I}_{n^{(s)}}^{(s)}\big)$ does not have saddle-points. Hence, the leading 
$|x_{k}-x_l|\cdot p_F \gg 1$, $k \ne l$,  asymptotic behaviour will be given by localizing the 
summand on the states belonging to $\ell_s$-critical classes, $s=1,\dots, r-1$. Upon inserting the local expression for the model's
form factors, a straightforward calculation  leads to the conclusion that  
\bem
C\big( \bs{x}_r; \bs{o}_r \big) \;  \simeq  \; 
 \sul{ \substack{ \bs{\ell}_{r-1}  \\ \in \mathbb{Z}^{r-1}  } }{} 
%
%
\bigg( \f{ 2\pi }{L} \bigg)^{\vartheta(\bs{\ell}_{r-1}, \bs{o}_r) } \cdot 
 \pl{s=1}{r-1} \bigg\{  \ex{ 2 i  \ell_s (x_{s+1}-x_s) p_F }  \bigg\}  \cdot 
\pl{s=1}{r}\Big\{  C^{(\ell_{s-1}; \ell_s)} \big( \nu_s^+, \nu_s^- \big)  \Big\}  \\
 \times \pl{s=1}{r} \Big\{ \mc{F}_{\mc{O}_s}(\ell_{s-1}, \ell_s)   \Big\}
 \; \cdot \;   
 \msc{S}_{\bs{\ell}_{r-1}}^{-}\Big( \big\{ \f{2\pi}{L}(x_{s+1}-x_s)   \big\}_1^{r-1}, \{\nu_s^-(\bs{\ell}_s)\}_1^r  \Big) 
\, \cdot  \, 
   \msc{S}_{\bs{\ell}_{r-1}}^{+}\Big(  \big\{ \f{2\pi}{L}(x_{s+1}-x_s) \big\}_1^{r-1}, \big\{\nu_s^+(\bs{\ell}_s) \big\}_1^r \Big)  \;. 
\label{ecriture asympt dominantes comme serie part trous}
\end{multline}
The summand at fixed $\bs{\ell}_{r-1}  \in  \mathbb{Z}^{r-1}$ is weighted by an algebraic factor 
in the volume $L$. The associated exponent $\vartheta(\bs{\ell}_{r-1}, \bs{o}_r)$ reads 
\beq
\vartheta(\bs{\ell}_{r-1}, \bs{o}_r) \; = \; \f{1}{2} \sul{s=1}{r} \Big\{ (\nu_s^+)^2 +  ( \nu_s^-)^2 \Big\}
\; - \; \sul{s=1}{r-1} \Big\{ \big( \nu_s^+ \,+\, \nu_s^- \, - \,  \nu_{s+1}^+ \, - \,  \nu_{s+1}^-  \big) \ell_s -2\ell_s^2 \Big\}  
\; - \; 2\sul{s=2}{r-1} \ell_s \ell_{s-1} \;. 
\enq
We do insist that each relative shift function $\nu_s$ does depend, \textit{a priori} 
on the integers $\ell_{s-1}, \ell_s$ labelling the critical class of the excited states
being connected by the $s^{\e{th}}$-operator.  

The form factor expansion has been recast with the help of multi-point restricted sums 
$\msc{S}_{\bs{\ell}_{r-1}}^{\pm}$. These refer to multi-dimensional sums that correspond to 
summing-up the whole set of contributions coming from excitations on Fermi surfaces,
for the intermediate states labelled by $s=1, \dots, r-1$, and belonging to given
critical classes labelled by the vector $\bs{\ell}_{r-1}$. 
The $+$, resp. the $-$, sum corresponds to the contributions stemming from the right, resp. left, Fermi
boundaries.  
 
The multi-point restricted sums of interest are given as the below multi-dimensional sums:
\beq
 \msc{S}_{ \bs{\ell}_{r-1} }^{\pm}\big( \{t_s\}_1^{r-1}, \{\nu_s\}_1^r \big) \;   = \; 
\pl{s=1}{r-1} \sul{ \substack{ n_p^{(s)}, n_h^{(s)} =0  \\ n_p^{(s)} - n_h^{(s)} = \pm \ell_s  } }{+\infty}  
\sul{ \mc{J}^{(s)}_{ n_p^{(s)}; n_h^{(s)} } }{ } 
\; \pl{s=1}{r-1} \mc{R}^{\pm} \Big( \mc{J}_{ n_p^{(s)}; n_h^{(s)} }^{(s)} \big| \nu_s, \nu_{s+1} ; t_s \Big)
 \pl{s=2}{r-1}  \varpi\Big( \mc{J}_{ n_p^{(s-1)}; n_h^{(s-1)} }^{(s-1)}; \mc{J}_{ n_p^{(s)}; n_h^{(s)} }^{(s)} \mid  \pm \nu_s\Big).
\label{definition sommes restreintes type plus et moins}
\enq
The summation in the above formula runs through all the possible choices of the sets of integers 
\beq
 \mc{J}_{ n_p^{(s)}; n_h^{(s)} }^{(s)} \; = \; \Big\{  \{p_a^{(s)}  \}_1^{n_p^{(s)}} \; ; \;  \{h_a^{(s)}\}_1^{n_h^{(s)}} \Big\} 
\enq
parametrizing the excitations on the relevant boundary (left for $-$ and right for $+$),
provided that these leads to an $\ell_s$-critical class. The summations are repeated 
for $s=1,\dots, r-1$. 
In more explicit terms, the formula corresponds to a multiple sum 
over particle-like $p_a^{(s)}$ and hole-like $h_a^{(s)}$
integers with the number $n_p^{(s)}$ of particle-like variables, resp.  $n_h^{(s)}$ of hole-like variables,
taking all admissible values in $\mathbb{N}$ that are compatible with the condition  $n_p^{(s)} - n_h^{(s)} = \pm \ell_s$.  
These particle and hole integers satisfy to the constraints 
\beq
 p_1^{(s)} <\dots < p_{n_p^{(s)}}^{(s)}  \;\;\e{with} \; \;   p_a^{(s)}  \in \mathbb{N}^{*}   
 \qquad \e{and} \qquad 
 h_1^{(s)}<\dots < h_{n_h^{(s)}}^{(s)}   \; \; \e{with} \; \;  h_a^{(s)}  \in \mathbb{N}^{*} \; . 
\enq
In what concerns the functions composing the summand,  $\varpi$  has been defined in \eqref{definition de la fonction  varpi} 
whereas the functions $\mc{R}^{\pm}$ read 
\bem
\mc{R}^{-}  \Big( \mc{J}_{n_p;n_h} \mid \nu, \eta ; t \Big)  =
\bigg(- \f{\sin[\pi \nu] }{ \pi } \cdot \f{\sin[\pi \eta] }{ \pi } \bigg)^{n_h} 
\cdot \f{ \pl{a<b}{n_p} (p_a-p_b)^2 \cdot \pl{a<b}{n_h} (h_a-h_b)^2 }
{ \pl{a=1}{n_p} \pl{b=1}{n_h} (p_a+h_b-1)^2 }  \\
			 \times \pl{a=1}{n_p} \bigg\{ \ex{it(1-p_a)} \Ga\pab{p_a+\nu,p_a-\eta }{p_a, p_a} \bigg\}
\cdot \pl{a=1}{n_h} \bigg\{ \ex{-ith_a } \Ga\pab{h_a-\nu, h_a+\eta}{h_a,h_a} \bigg\}  \nonumber
\end{multline}
and
\bem
\mc{R}^{+}  \Big( \mc{J}_{n_p;n_h} \mid \nu, \eta ; t  \Big)  =
\bigg(- \f{\sin[\pi \nu] }{ \pi } \cdot \f{\sin[\pi \eta] }{ \pi } \bigg)^{n_h} 
				 \cdot \f{ \pl{a<b}{n_p} (p_a-p_b)^2 \cdot \pl{a<b}{n_h} (h_a-h_b)^2 }
{ \pl{a=1}{n_p} \pl{b=1}{n_h} (p_a+h_b-1)^2 }   \\
\times  \pl{a=1}{n_p} \bigg\{ \ex{itp_a} \Ga\pab{p_a-\nu, p_a + \eta }{p_a, p_a} \bigg\}
\cdot \pl{a=1}{n_h} \bigg\{ \ex{it(h_a-1) } \Ga\pab{h_a+\nu, h_a - \eta}{h_a, h_a} \bigg\}  \;. 
\nonumber
\end{multline}
Above, the set $\mc{J}_{n_p;n_h}$ is as given in \eqref{definition ensembles part trous modeles}. 
Quite remarkably, the multi-particle restricted sums can be resummed in a quite compact way, generalising the results of 
\cite{KozKitMailSlaTerRestrictedSums,OlshanskiPointProcessAndInfiniteSymmetricGroup}
\bem
\msc{S}_{\bs{\ell}_{r-1}}^{\pm}\big( \{t_s\}_1^{r-1}, \{\nu_s\}_1^r \big)  \; = \;  
\pl{s=1}{r-1} \Bigg\{  \ex{ \pm i  t_s \f{ \ell_s(\ell_s+1) }{2} }  
 G\bigg( \ba{cc}   1\pm(\ell_s-\nu_s) , 1 \pm (\ell_s + \nu_{s+1})  \\  
						1\mp\nu_s , 1\pm\nu_{s+1} \ea \bigg)   \Bigg\} \\
\times 						
\pl{s=2}{r-1} G\bigg( \ba{cc}   1 \pm \nu_s , 1 \pm( \ell_{s-1}-\ell_s + \nu_s)  \\  
						1\mp(\ell_s - \nu_s) , 1 \pm (\ell_{s-1} + \nu_{s})  \ea \bigg)
\; \cdot \; \pl{b>a}{r} \Big( 1-\ex{ \pm i \sul{s=a}{b-1}t_s}  \Big)^{(\nu_a + \kappa_a)(\nu_b + \kappa_b)		} \;. 
\label{formule calcul exact somme restreinte multiples}
\end{multline}
The product appearing at the bottom right of \eqref{formule calcul exact somme restreinte multiples}
involves integers $\kappa_a$, $a=1,\dots,s$ which are defined in terms of the integers $\ell_a$, $a=1,\dots, s-1$, as
\beq
\kappa_s \; = \; \ell_{s-1}-\ell_{s} \quad \e{for} \quad s= 1, \dots, r \;
 \qquad \e{so} \; \e{that} \qquad \sul{a=1}{r} \kappa_a \; = \; 0
\label{ecriture parametrization kappa1}
\enq
where, as above, we agree upon $\ell_0 = \ell_r = 0$. 

In the present paper we do not pretend to prove the above summation formulae. 
We however give a formal derivation. These arguments does not 
lead, however, to a proof in that several justifications relative to the exchangeability of symbols
and limits are omitted. We leave the rigorous proof of the summation formulae
to some subsequence publication. Our present formal justification is based on the use
of an auxiliary, "model", form factor series. The latter can be recast in two ways,
\begin{itemize}
\item as a Toeplitz determinant generated by a symbol with Fisher-Hartwig singularities, this by generalizing the 
techniques of form factor summations developed in \cite{KorepinSlavnovTimeDepCorrImpBoseGas}; 
\item as an asymptotic series whose individual summands directly involve multi-particle 
restricted sums. 
\end{itemize}
Comparisons of the large-$N$ limit of both expressions yields 
the sought identities. We refer the reader to Appendix \ref{Appendix summation identity}
for our formal proof.

Building on the above results allowing one to sum up the multi-point restricted sums and
taking into account their most natural parametrization in terms of the $ \kappa_a$
integers, we change the summation variables to the more convenient $\kappa$-like reparametrisation: 
\beq
\ell_s(\bs{\kappa}_r) \; = \; \sul{a=s+1}{r} \kappa_a  \qquad  \e{for} \qquad s=1,\dots, r-1 \;.  
\label{ecriture parametrization ls et kappa s}
\enq
It is then readily checked that, for any set of parameters $\{t_a\}$ and $\{ \nu_a \}$,
\beq
 \sul{s=1}{r-1} \ell_s(\bs{\kappa}_r) \cdot  t_s \; = \; \sul{s=2}{r} \ov{\bs{t}}_{s-1} \kappa_s ,
\label{ecriture identite resommation exposant oscillant ells vers vars kappas}
\enq
and 
\beq
2\sul{s=1}{r} [\ell_s(\bs{\kappa}_r)]^2  \; - \;  
2 \sul{s=2}{r-1}\ell_s(\bs{\kappa}_r)  \ell_{s-1}(\bs{\kappa}_r) \;  + \; 
 2 \sul{s=1}{r-1}\Big\{ (\nu_{s+1}-\nu_s)\ell_s(\bs{\kappa}_r) \Big\} 
\; + \; \sul{s=1}{r} \nu_s^2  \; = \; \sul{s=1}{r} (\nu_s + \kappa_s)^2 \; ,
\label{ecriture resomation somme complexe ells vers somme simple ac kappas}
\enq
where in \eqref{ecriture identite resommation exposant oscillant ells vers vars kappas} we have used a similar notation as in \eqref{definition Ns et os sum}.
Upon using the explicit expressions for the multi-particle restricted sums, 
we are thus led to 
\bem
C\big( \bs{x}_r; \bs{o}_r \big) \;  = \; 
\sul{ \substack{ \bs{\kappa}_{r}   \in \mathbb{Z}^{r}  \\   \sul{}{}\kappa_a = 0  } }{} 
\pl{s=1}{r} \bigg\{ \ex{ 2 i   p_F \kappa_s x_{s} }  \bigg\}   
\cdot \mc{F}\Big( \{ \kappa_a \}_1^{r} ; \{ o_a \}_1^r   \Big) \\
\times \ \pl{s=1}{r}
 \bigg\{ \bigg( \f{ 2\pi }{L} \bigg)^{ \f{1}{2} [\th_s^+(\kappa_s)  ]^2 + \f{1}{2} [ \th_s^-(\kappa_s) ]^2}  
\bigg\} \cdot 
\pl{ b>a }{ r } \Bigg\{  \Big[ 1- \ex{\f{ 2i\pi}{L} (x_b-x_a) } \Big]^{\th_b^+(\kappa_b) \th_a^+(\kappa_a)} \cdot 
\Big[ 1- \ex{-\f{ 2i\pi}{L} (x_b-x_a) } \Big]^{\th_b^-(\kappa_b) \th_a^-(\kappa_a)} \Bigg\} \;. 
\label{ecriture forme resommee asympt serie FF}
\end{multline}
There, we have set
\beq
\th_b^{\pm}(\kappa_b) \; = \; \nu_b^{\pm} + \kappa_b \;. 
\enq
We do stress that $\th_b^{\pm}$ has an explicit dependence on $\kappa_b$ but also an \textit{implicit}
one through the relative shift function $\nu_b^{\pm}$. Furthermore, we have made use of the fact 
that the relative shift functions $\nu_s^{\pm}$ \textit{only}
depend on the parameter $\kappa_s$. Finally, the asymptotic formula 
involves the factor 
\beq
\mc{F}\Big( \{ \kappa_a \}_1^{r} ; \{ o_a \}_1^r   \Big)   \; = \;  
\pl{s=1}{r} \mc{F}_{\mc{O}_s}\big(\ell_{s-1}(\bs{\kappa}_r), \ell_s(\bs{\kappa}_r) \big)   
%
%
\;. 
\enq
The above amplitude has a crystal clear interpretation: the pre-factor in front of the power-law decay of the 
multi-point correlation function associated with an excitation belonging to the $\bs{\ell}_{r-1}(\bs{\kappa}_r)$-critical classes is precisely given by the product of form factors of local operators 
taken between the typical representatives of the $\bs{\ell}_{r-1}(\bs{\kappa}_r)$-critical classes of interest. Note that, for a given $ \bs{\kappa}_r $, the form of the $L$-dependence 
in \eqref{ecriture forme resommee asympt serie FF}  is already reminiscent of the expression 
for multi-point correlation functions in a conformal invariant theory of a strip of width $L$ \cite{CardyConformalExponents}.

It is straightforward to take the $L\tend +\infty$ limit in \eqref{ecriture forme resommee asympt serie FF}. This leads to 
\beq
C\big( \bs{x}_r; \bs{o}_r \big) \;  = \; 
\sul{ \substack{ \bs{\kappa}_{r}   \in \mathbb{Z}^{r}  \\   \sul{}{}\kappa_a = 0  } }{} 
\pl{s=1}{r} \bigg\{ \ex{ 2 i   p_F \kappa_s x_{s} }  \bigg\}   
\cdot \mc{F}\Big( \{ \kappa_a \}_1^{r} ; \{ o_a \}_1^r   \Big) 
\cdot \pl{ b>a }{ r } \Bigg\{  \big[ i(x_b-x_a) \big]^{\th_b^-(\kappa_b) \th_a^-(\kappa_a)} \cdot 
\big[ -i(x_b-x_a) \big]^{\th_b^+(\kappa_b) \th_a^+(\kappa_a)}  \Bigg\} \;,
\label{ecriture formule DA asymptotique multi pts avec espacement arbitraire}
\enq
where we have used that
\beq
\sul{s=1}{r} \th_s^{\pm}(\kappa_s) \; = \; 0 \;. 
\enq
Note that the above  asymptotic expansion provides one with an expression that is symmetric under a simultaneous 
permutation 
\beq
\big(\bs{x}_r , \bs{o}_r \big) \; \mapsto \; \big(\bs{x}_r^{\sg} , \bs{o}_r^{\sg} \big)
\quad \e{with} \quad \bs{x}_r^{\sg} \; = \; \big(x_{\sg(1)}, \dots ,x_{\sg(r)} \big) \qquad \sg \in \mf{S}_r \;. 
\enq
This translates the fact that the local operators $\mc{O}_r(x_r)$ commute at different distances
and, in particular, in the long-distance regime. 




\section{Applications}
\label{Section applications to integrable models}

\subsection{The conformal regime}

The asymptotic formula \eqref{ecriture formule DA asymptotique multi pts avec espacement arbitraire} is quite general 
in that it does not assume any specific order of magnitude for the spacing between the various "space"-parameters
$x_a$. The sole constraint is that all have to be pairwise large, \textit{viz}.
\beq
|x_k - x_{\ell} |\cdot p_F \; \gg \; 1 \; \qquad \e{for} \qquad k \not= \ell \;. 
\enq
Thus, in such a general setting, determining the leading term of the asymptotics \eqref{ecriture formule DA asymptotique multi pts avec espacement arbitraire} leads to a very complex minimisation problem. However, the situation gets much simpler in the 
so-called conformal regime of the correlation functions. In the latter case, all distances scale with the same magnitude $R$, 
\beq
x_k \; = \; R \cdot  z_k \qquad   0<\eps < |z_k - z_{\ell} | < \eps^{-1} \; \qquad \e{for} \qquad k \not= \ell \quad \e{and} \;\e{some}
\; \; \eps >0\;. 
\enq
One then has 
\bem
C\big( R \cdot \bs{z}_r; \bs{o}_r \big) \;  \simeq  \; 
\sul{ \substack{ \bs{\kappa}_{r}   \in \mathbb{Z}^{r}  \\   \sul{}{}\kappa_a = 0  } }{} 
 \pl{s=1}{r} \bigg\{ \bigg( \f{1}{R} \bigg)^{ \frac{[\th_s^+(\kappa_s)]^2}{2} + \frac{[\th_s^-(\kappa_s)]^2}{2} } \bigg\}
 \pl{s=1}{r} \bigg\{ \ex{ 2 i   p_F R \kappa_s  z_{s} }  \bigg\}   
\cdot \mc{F}\Big( \{ \kappa_a \}_1^{r} ; \{ o_a \}_1^r   \Big) \\
\times \pl{ b>a }{ r } \Bigg\{  \big[ i(z_b-z_a) \big]^{\th_b^-(\kappa_b) \th_a^-(\kappa_a)} \cdot 
\big[ -i(z_b-z_a) \big]^{\th_b^+(\kappa_b) \th_a^+(\kappa_a)}  \Bigg\} 
\end{multline}
using again,
\beq
\sul{s=1}{r} \th_s^{\pm}(\kappa_s) \; = \; 0 \;. 
\enq

The leading asymptotics are then obtained by choosing an integer vector  $ \bs{\kappa}_r \in \mathbb{Z}^r$
realising the minimum of 
\beq
\bs{\kappa}_r  \; \mapsto \;  \sul{s=1}{r} [\th_s^+(\kappa_s)]^2 + [\th_s^-(\kappa_s)]^2 \;. 
\enq

In practical situations (such as Bethe Ansatz solvable models, \textit{cf}. later on)
the existence and uniqueness of the minimum, for generic values of the coupling constants,
follows by mimicking the reasonings presented in \cite{EhrhardtAsymptoticBehaviorOfFischerHartwigToeplitzGeneralCase}.

\subsection{Bethe Ansatz solvable models}

In this subsection we specialise our results to the case of two Bethe Ansatz solvable models,
the XXZ spin-$1/2$ chain and the non-linear Schr\"{o}dinger model. The matter is that 
one has quite explicit expressions for the shift functions arising in the description of these
models in terms of solutions to linear integral equations. The latter thus give access to the critical exponents.

The XXZ spin-$1/2$ chain corresponds to the Hamiltonian
 \beq
\bs{H}_{XXZ} \; = \; \sum_{k=1}^{L}\left(
 \sigma^x_{k}\sigma^x_{k+1}+\sigma^y_{k}\sigma^y_{k+1}
 +\Delta(\sigma^z_{k}\sigma^z_{k+1}-1)\right)-\frac{h}{2}\sum_{k=1}^{L}\sigma^z_{k}.
 \enq
Here $\sigma^{x,y,z}_{k}$ are the spin operators (Pauli matrices) acting on the $k$-th site
of the chain,  $h$ is an external magnetic field and the model is subject to periodic boundary conditions. 
The XXZ spin chain above exhibits different phases depending on the value of  the anisotropy parameter  $\Delta$. 
We only focus on the critical regime $-1<\De<1$ where we set $\Delta=\cos \zeta$.  
It is known that, in its massless phase, the excitations in the XXZ-chain can be either
of bound state nature (so-called string solutions) or particle-hole one. As argued previously,
we solely focus on the particle-hole part of the spectrum.

The non-linear Schr\"{o}dinger model corresponds to the Hamiltonian
\beq
\bs{  H} _{NLS} =\Int{0}{L} \paa{ \Dp{y} \Phi^{\dagger}\!\pa{y} \Dp{y} \Phi\pa{y} + c \, \Phi^{\dagger}\!\pa{y}\Phi^{\dagger}\!\pa{y} \Phi\pa{y} \Phi\pa{y}
\, - \,  \mu \, \Phi^{\dagger}\!\pa{y} \Phi\pa{y} } \dd y\;.
\label{definition Hamiltonien Bose}
\enq
The model is defined on a circle of length $L$, so that the canonical Bose fields $\Phi$, $\Phi^{\dagger}$ are subject to $L$-periodic 
boundary conditions. We solely focus on the repulsive regime $c>0$ in the presence of a positive chemical potential $\mu > 0$.
In this model, one can show that the excitations are only given by particles and holes.

One can show using Bethe Ansatz methods that, for both models, the general shift functions -- in the sense of \eqref{ecriture definition fonction comptage} -- take the form 
\beq
F_{\mc{R}_n;s}(\la) \; = \; \; -\; \ov{\bs{o}}_s \Big( \f{ Z(\la) }{ 2 } \; + \; \upsilon \phi(\la,q)  \Big)
\; - \; \upsilon \sul{a=1}{n} \Big[ \phi\big(\la,\mu_{p_a} \big) \, - \,  \phi\big(\la,\mu_{h_a} \big)   \Big]\, ,
\enq
in which the rapidities are the unique solutions to 
\beq
\xi( \mu_a) \; = \; \f{ a }{ L }  \qquad \e{with} \qquad \xi(\om) \; = \; \f{ p(\om) }{2\pi} \; + \; \f{D}{2} \;, 
\enq
with $p$ representing the dressed momentum associated with the model and $\ov{\bs{o}}_s$ has been defined in \eqref{definition Ns et os sum}.
Although the form for $\xi$ and $F_{\mc{R}_n;s}$ is similar for both models, the definition of the functions arising in their expressions differ.  Also, one should set 
\begin{itemize}
\item  $\upsilon = 1$ for the non-linear Schr\"{o}dinger model 
\item $\upsilon =-1$ for the XXZ chain. 
\end{itemize}
The function $\phi$ is the dressed phase and $Z$ the dressed charge. Setting 
\beq
\th(\la) \; = \; i \ln \Big( \f{ ic + \la }{ i c - \la } \Big) \quad \e{for} \quad NLSM \qquad \e{and} \qquad
\th(\la) \; = \; i \ln \bigg( \f{ \sinh(i\zeta + \la) }{ \sinh(i\zeta - \la)  } \bigg) \quad \e{for} \quad XXZ \;, 
\enq
one has that $p$ solves the integro-differential equation under the requirement that $p(\la)=-p(-\la)$:
\beq
p(\la) \; - \; \upsilon \Int{-q}{q} \th(\la-\mu) \cdot p^{\prime}(\mu) \cdot \f{ \dd \mu }{2\pi} \; = \; p_{0}(\la) \quad
\e{with} \qquad
p_0(\la) \; = \;  \left\{ \ba{cc}  \la & \e{for} \; \e{NLSM}   \vspace{2mm} \\  
						 i \ln \bigg( \f{ \sinh(i\zeta/2 + \la) }{ \sinh(i\zeta/2 - \la)  } \bigg)  & \e{for} \; \e{XXZ}   \ea \right.   \;. 
\enq
The functions $Z$ and $\phi$ solve the Lieb integral equations
\beq
Z(\la) \; - \; \upsilon \Int{-q}{q} \th^{\prime}(\la-\mu) Z(\mu) \cdot \f{ \dd \mu }{ 2\pi } \; = \; 1
\qquad 
\phi(\la,\nu) \; - \; \upsilon \Int{-q}{q} \th^{\prime}(\la-\mu) \phi(\mu,\nu) \cdot \f{ \dd \mu }{ 2\pi } \; = \; 
\f{ \th(\la-\nu) }{ 2\pi } \;. 
\enq

As a consequence, the relative shift function $\nu_{s}$ between critical excited states belonging to the 
$\ell_s$ and $\ell_{s-1}$ classes and having a change in the pseudo-particle number of $o_s$ takes the form 
\beq
\nu_s(\la) \; = \; o_s \Big( \f{ Z(\la) }{ 2 } \; + \; \upsilon \phi(\la,q)  \Big)
\; + \; (\ell_{s-1} -\ell_s ) \Big( Z(\la) - 1  \Big) \;. 
\enq

In particular, then, the critical exponents takes the explicit expressions 
\beq
\th^{\pm}_s(\kappa) \; = \; \kappa Z(q) \;   \mp \;  \f{o_s}{2} Z^{-1}(q) \;. 
\enq

These data are already enough so as to extract, in the large-distance conformal limit, the
asymptotic behaviour of multi-point correlation functions of the XXZ spin-$\tf{1}{2}$ chain. 
We shall discuss the example of the correlator
\beq
C_{xxxx} \; = \; \bra{\Psi_g} \sg_{x_1}^{x} \cdot \sg_{x_2}^{x} \cdot \sg_{x_3}^{x} \cdot \sg_{x_4}^{x} \ket{\Psi_g} \;. 
\enq

In order to apply the method developed in the present paper, one ought to decompose the operators $\sg_{x_a}^{x}$
onto operators which change the particle number in a definite way, namely the $\sg_{x_a}^{\pm}$ operators. 
Due to the conservation of the longitudinal total spin, such a decomposition leads to 
\beq
C_{xxxx} \; = \; \sul{ \substack{ \eps_a = \pm \\  \sum \eps_a = 0 } }{} 
\bra{\Psi_g} \sg_{x_1}^{\eps_1} \cdot \sg_{x_2}^{\eps_2} \cdot \sg_{x_3}^{\eps_3} \cdot \sg_{x_4}^{\eps_4} \ket{\Psi_g} \;. 
\enq

We shall consider the large-$x$ asymptotic behaviour of $C_{xxxx}$ in the conformal scaling regime 
where $x_a = R \cdot z^{\prime}_a$, $R$ is the large parameter and the $z^{\prime}_a$ are fixed 
and pairwise distinct. Then, it follows from the previous calculations that 
\bem
\bra{\Psi_g} \sg_{x_1}^{\eps_1} \cdot \sg_{x_2}^{\eps_2} \cdot \sg_{x_3}^{\eps_3} \cdot \sg_{x_4}^{\eps_4} \ket{\Psi_g}
\; = \; \sul{ \substack{ \bs{\kappa}_{4}   \in \mathbb{Z}^{4}  \\   \sul{}{}\kappa_a = 0  } }{} 
 \pl{s=1}{4} \bigg\{ \bigg( \f{1}{R} \bigg)^{ \frac{ [\vartheta_{\eps_s}(\kappa_s)]^2 }{ 2 } + \frac{ [\vartheta_{-\eps_s}(\kappa_s)]^2 }{2} } \bigg\}
 \pl{s=1}{4} \bigg\{ \ex{ 2 i   p_F R \kappa_s  z_{s} }  \bigg\}   
\cdot \mc{F}\Big( \{ \kappa_a \}_1^{r} ; \{ \eps_a \}_1^r   \Big) \\
\times \pl{ b>a }{ r } \Bigg\{  \big[ i(z_b-z_a) \big]^{\vartheta_{-\eps_b}(\kappa_b) \vartheta_{-\eps_a}(\kappa_a)} \cdot 
\big[ -i(z_b-z_a) \big]^{\vartheta_{\eps_b}(\kappa_b) \vartheta_{\eps_a}(\kappa_a)}  \Bigg\} \;. 
\end{multline}
Above $\mc{F}\Big( \{ \kappa_a \}_1^{r} ; \{ \eps_a \}_1^r   \Big)$ represents the properly normalized in the 
volume $L$ product of form factors of operators $\sg^{\eps_a}$. Also, we agree upon 
\beq
\vartheta_{\eps}(\kappa) \; = \; \kappa Z(q) \,- \, \f{ \eps  }{2 Z(q) } \;. 
\enq
In order to access to the leading asymptotics in the distance
one should choose the combination of integers $\bs{\kappa}_{4}=(\kappa_1,\dots, \kappa_4)$
that minimises the exponent of $R$. In fact, independently of the operator (\textit{ie} the sign of $\eps_a$)
one has
\beq
[\vartheta_{\eps_s}(\kappa_s)]^2 + [\vartheta_{-\eps_s}(\kappa_s)]^2 \; = \;
2 \big( \kappa_s Z(q) \big)^2 \, + \, \f{  1 }{2 Z^2 (q) } \;. 
\enq
As a consequence, independently of the choice of the $\eps_a$'s, the leading asymptotics are given by the
choice $\bs{\kappa}_4 = 0$. It is then straightforward to see that 
\beq
C_{xxxx} \; = \; 2 \mc{F}_{\sg^+}^{2}\big(0,0) \,   \mc{F}_{\sg^-}^{2}\big(0,0) \cdot \bigg\{   
\Big| \f{  (x_2-x_1)\cdot (x_4-x_3)  }{  (x_3-x_1)   \cdot (x_4-x_1)\cdot (x_3-x_2) \cdot (x_4-x_2)   } \Big|^{   \f{  1 }{2 Z^2 (q) } } 
\; + \; (2 \leftrightarrow 3) \; + \; (2 \leftrightarrow 4)  \bigg\} \; +  \; \dots \;. 
\enq
Above, $\mc{F}_{\sg^{\pm}}\big(0,0)$ are the aforediscussed form factors of the $\sg^{\pm}$
operators between appropriate states of the $0$-critical class.

Although we have focused in the present example on ultra-local elementary operators acting on a single site, 
our approach would works just as good for higher composite operators such as finite products of local operators on adjacent site,
such as 
\beq
\mc{O}(x) \; = \; \pl{a=1}{p} \sg_{x+a}^{\eps_a} \;. 
\enq

\section*{Conclusion}

In this paper, we have generalized to the multi-point case the restricted sum formalism developed in  
\cite{KozKitMailSlaTerRestrictedSums}
for  the study of the form factor expansion of two-point correlation functions in massless models. 
In this framework, the computation of the form factor series, for large but finite volume and in the large distance regime, reduces to the computation of multi-dimensional sums over particular classes of low-energy excited states.
Our formalism naturally applies to quantum integrable models such as the XXZ spin-1/2 chain in the massless regime or the quantum non-linear Schr\"odinger model in the repulsive regime. It also applies, upon quite reasonable assumptions
on the way of parametrizing the model's spectrum and the structure of the model's form factors,
to more general massless one-dimensional quantum models. Let us finally mention that this approach to multi-point correlation functions, developed here in the static case, can be extended, as in \cite{KozKitMailSlaTerRestrictedSumsEdgeAndLongTime},  to the time-dependent case, {\it ie} to the study of the large-distance and long-time asymptotic behaviour of multi-point dynamical correlation functions.


\section*{Acknowledgements}

K.K.K., J.M.M. and V. T. are supported by CNRS. 
N.K, K.K.K., J.M.M. and V.T. are supported by ANR grant ``DIADEMS''.
This work has been partly done within the 
financing of the grant . 
K.K.K. acknowledges support from the 
the Burgundy region PARI 2013 FABER grant "Structures et asymptotiques d'int\'{e}grales multiples"
and PEPS-PTI "Asymptotique d'int\'{e}grales multiples" grant. V. T. would like to thank LPTHE, University Paris VI, for hospitality.





\appendix

\section{The summation identity}
\label{Appendix summation identity}

 In order to establish the summation identities of interest, we shall 
 compute the large-$N$ behaviour of the sum 
\beq
\mc{S}_N^{(r)} \; = \;  \pl{k=1}{r-1}\bigg\{ \sul{ \substack{ \ell_a^{(k)} \in \mathbb{Z} \\   \ell_1^{(k)} < \dots < \ell_N^{(k)} } }{}   \bigg\}
\cdot \bigg( \f{ \sin[\pi \nu_r]  }{ \pi }  \bigg)^N \cdot 
\pl{s=1}{r-1}\pl{a=1}{N} \Bigg\{  \f{ \sin[\pi \nu_s]  }{ \pi }  \exp\Big\{ i t_s \la_{\ell_a^{(s)}}^{(s)}  \Big\} \Bigg\} \cdot 
\pl{k=1}{r} \bigg\{ \det_N\bigg[ \f{1}{ \la_{\ell_a^{(k-1)}}^{(k-1)} \;  - \;  \la_{\ell_b^{(k)}}^{(k)}  }\bigg]  \bigg\} 
\label{ecriture somme restr}
\enq
in two ways. 

The definition of $\mc{S}^{(r)}_N$ involves the numbers $\la_a^{(k)}$ defined as
\beq
\la_a^{(0)} = a \; - \; \f{N+1}{2}   \qquad \e{and} \qquad  
\la_{a}^{(k)} = a - \f{N+1}{2} \, - \, \ov{\bs{\nu}}_k \;  \qquad \e{for} \quad k=1, \dots , r \;.  
\enq
Above, we have adopted the convenient shorthand notation: for a vector $\bs{\eta}_k$ we denote
\beq
\ov{\bs{\eta}}_k \; = \; \sul{s=1}{k}\eta_s  \;. 
\enq
Finally, in \eqref{ecriture somme restr}, $\nu_s$ and $t_s$ are some auxiliary complex numbers such that 
\beq
 \Re(\nu_s) \not \in \f{1}{2} \, +  \, \mathbb{Z}  \qquad \e{and} \qquad 
 \ov{\bs{t}}_s\in \intff{0}{2\pi}, \quad s=1,\dots, r-1\;. 
\enq

Furthermore, the two boundary sequences of integers read
\beq
\ell^{(0)}_a \; = \; \ell^{(r)}_a \; = \; a  \;. 
\enq

The $(r-1) \times N$-fold sums \eqref{ecriture somme restr} defining $\mc{S}_N^{(r)}$ can be recast in two different ways. On the one hand, one can relate it to
a Toeplitz matrix. On the other hand, one can represent it as a sum over particle-hole like excitations
associated with each intermediate state arising in the expansion. The latter 
generates the multidimensional restricted sums \eqref{definition sommes restreintes type plus et moins}
that arise in the context of form factor expansions of multi-point correlation functions.

\subsection{Toeplitz determinant representation}

Let $\mc{S}_{N; \bs{\ell}^{(n)}_N }^{(n)}$ satisfy the induction 
\beq
\mc{S}_{N; \bs{\ell}^{(n+1)}_N}^{(n+1)} \; = \; 
 \sul{ \substack{ \ell_a^{(n)} \in \mathbb{Z} \\   \ell_1^{(n)} < \dots < \ell_N^{(n)} } }{}   
\pl{a=1}{N} \Bigg\{  \f{ \sin[\pi \nu_{n+1}]  }{ \pi }  \ex{i t_n \big( \ell_a^{(n)} - \f{N+1}{2} - \ov{\bs{\nu}}_n \big) }\Bigg\} \cdot 
 \det_N\bigg[ \f{1}{ \ell_b^{(n)} \;  - \;  \ell_a^{(n+1)} + \nu_{n+1} }\bigg]  \bigg\} \cdot \mc{S}_{N; \bs{\ell}^{(n)}_N }^{(n)}  \;, 
\enq
and be subject to the initialization condition
\beq
\mc{S}_{N; \bs{\ell}^{(1)}_N }^{(1)} \; = \; \bigg( \f{ \sin[\pi \nu_1]  }{ \pi }  \bigg)^N 
\cdot \det_N\bigg[ \f{1}{ b - \ell_a^{(1)}   +  \nu_1  }\bigg]   \;. 
\enq
It is then readily seen that with $\bs{\ell}_N^{(r)}=(1,\dots,N)$
\beq
\mc{S}_{N; \bs{\ell}^{(r)}_N }^{(r)} \; = \;  \mc{S}_{N }^{(r)} \;. 
\enq
Furthermore, we shall establish by induction on $n$ that 
\beq
\mc{S}_{N; \bs{\ell}^{(n+1)}_N }^{(n+1)} \; = \;  \pl{s=1}{n} \Big\{  \ex{-iN \ov{\bs{\nu}}_s t_s}  \Big\} \cdot 
\pl{a=1}{N} \Big\{  \ex{i\ov{\bs{t}}_n\big( \ell_a^{(n+1)}-\f{N+1}{2}  \big) }   \Big\}  
\cdot \det_N\Big[ c_{\ell_a^{(n+1)}-b}\big[ \, \chi_{n+1}  \big]  \Big]  \;.  
\enq
Above, given a function $f$ on $\intff{0}{2 \pi}$, we have set 
\beq
c_k[f] \; = \; \Int{0}{2\pi}  \ex{-ik \th} f(\th) \cdot \f{ \dd \th }{ 2\pi } \;. 
\enq
The function $\chi_n$ is expressed in terms of elementary jump-like Fisher--Hartwig symbols
\beq
\chi_n \;  = \; \chi_{ \nu_1 , 0} \cdot \pl{s=1}{n-1} \chi_{ \nu_{s+1} , \ov{\bs{t}}_s}
\enq
where 
\beq
\chi_{\de, \vp}(\th) \; = \; \ex{i(\th-\vp+\pi)\de } \Big\{ \, \bs{1}_{\intfo{0}{\vp}} \; + \; 
		\ex{-2i\pi\de }   \bs{1}_{ \intfo{\vp}{2\pi} }	\, \Big\}  \;. 
\enq
One can recast $\mc{S}_{N; \bs{\ell}^{(1)}_N }^{(1)}$ in terms of a Toeplitz determinant. Indeed, one has
\beq
c_{j}\big[\chi_{ \nu_1 , 0}\big] \; = \; \ex{-i\pi \nu_1 }\Int{0}{2\pi} \ex{i(\nu_1-j) \th }  \cdot \f{\dd \th }{2\pi }
\; = \; \f{ \ex{i\pi \nu_1}  - \ex{-i\pi \nu_1} }{ i2\pi (\nu_1 - j )  } \; = \; \f{ \sin[\pi \nu_1] }{ \pi (\nu_1 - j ) } \;. 
\enq
Thus, all in all, 
\beq
\mc{S}_{N; \bs{\ell}^{(1)}_N }^{(1)} \; = \;  \det_{N}\Big[ c_{\ell_a^{(1)} - b}\big[\chi_{ \nu_1 ,0 }\big]  \Big] \;, 
\enq
so that the induction hypothesis does indeed hold for $n=1$. Then, assume it holds for some $n$. 
The antisymmetry of the determinants allows one to replace the summation over a fundamental symplex by one 
over the whole of $\mathbb{Z}^N$ normalized by $\tf{1}{N!}$. Then, using the antisymmetry of the determinants, one 
can replace one of the determinants by $N!$ times the product of its diagonal entries. 
This ultimately yields:
\bem
\mc{S}_{N; \bs{\ell}^{(n+1)}_N }^{(n+1)} \; = \;  \pl{s=1}{n} \Big\{  \ex{-i N \ov{\bs{\nu}}_s t_s}  \Big\} \cdot 
\pl{a=1}{N} \Big\{  \ex{i\ov{\bs{t}}_n\big( \ell_a^{(n+1)}-\f{N+1}{2}  \big) }   \Big\}   \\
\times  \sul{ \bs{\ell}^{(n)}_N \in \mathbb{Z}^N  }{}   
\pl{a=1}{N} \Bigg\{  \f{ \sin[\pi \nu_{n+1}]  }{ \pi }  \ex{i \ov{\bs{t}}_n \big( \ell_a^{(n)} - \ell_a^{(n+1)}  \big) }\Bigg\} \cdot 
 \det_N\bigg[ \f{ c_{\ell_b^{(n)}-b}\big[ \chi_{n}  \big] }{ \ell_b^{(n)} \;  - \;  \ell_a^{(n+1)} + \nu_{n+1} }\bigg]  \;. 
\end{multline}
Entering the sums into the lines of the determinant, one gets the representation:
\beq
\mc{S}_{N; \bs{\ell}^{(n+1)}_N }^{(n+1)} \; = \; \pl{s=1}{n} \Big\{  \ex{-i N \ov{\bs{\nu}}_s t_s}  \Big\} \cdot 
\pl{a=1}{N} \Big\{  \ex{i\ov{\bs{t}}_n\big( \ell_a^{(n+1)}-\f{N+1}{2}  \big) }   \Big\} 
   \cdot  \det_N\Big[ M^{(n)}_{\ell_a^{(n+1)}b} \Big]  
\enq
where we have set
\bem
M^{(n)}_{ a b} \; = \; 
 \sul{ \ell \in \mathbb{Z}  }{}   
 \f{ \sin[\pi \nu_{n+1}]  }{ \pi }  \cdot   \ex{i \ov{\bs{t}}_n ( \ell -a  ) } \cdot 
\f{ c_{\ell-b}\big[ \chi_{n}  \big] }{ \ell   -   a + \nu_{n+1} }  
\; = \; \f{ \sin[\pi \nu_{n+1}]  }{ \pi }  \cdot  \ex{ - i \ov{\bs{t}}_n  a } \Int{0}{2\pi} \ex{i b \th } \chi_n(\th)  
\sul{\ell \in \mathbb{Z}}{} \f{ \ex{i \ell \big( \ov{\bs{t}}_n - \th \big) } }{ \ell   -  a + \nu_{n+1} } \cdot \f{ \dd \th }{ 2\pi } \;. 
\nonumber
\end{multline}
Using that, for $0< t < 2\pi$ 
\beq
\sul{\ell \in \mathbb{Z} }{}  \f{ \ex{it \ell} }{ \ell + a } \; = \; \f{ 2i\pi  \ex{-ia t }  }{  1-\ex{-2i\pi a} } \;, 
\enq
one gets 
\beq
\sul{\ell \in \mathbb{Z}}{} \f{ \ex{i \ell \big( \ov{\bs{t}}_n - \th \big) } }{ \ell   -  a + \nu_{n+1} }  \; = \; 
\f{ \pi \ex{i\pi (\nu_{n+1}-a) } }{ \sin[\pi(\nu_{n+1}-a)] } \cdot  \ex{ - i (\nu_{n+1}-a) \big( \ov{\bs{t}}_n - \th \big)} 
\cdot \Big\{ \bs{1}_{\intfo{0}{\ov{\bs{t}}_n}}(\th) \; + \; \ex{- 2 i \pi \nu_{n+1} }  \bs{1}_{\intfo{\ov{\bs{t}}_n}{2\pi}}(\th)  \Big\}
\enq
leading to 
\beq
M^{(n)}_{ a b} \; = \;  c_{a-b}\big[ \chi_{n+1}  \big]\, ,
\enq
hence establishing the induction hypothesis at $n+1$. 
Then, the large-$N$ asymptotic behavior of $ \mc{S}_N^{(r)}$ can be obtained, say, from the results established in  \cite{EhrhardtAsymptoticBehaviorOfFischerHartwigToeplitzGeneralCase}:
\bem
\mc{S}_N^{(r)} \; = \; \pl{s=1}{r-1} \Big\{ \ex{-iN t_s \bs{\ov{\nu}}_{s} } \cdot  \ex{i N  \kappa_{s+1} \ov{\bs{t}}_{s} }  \Big\}
\cdot  \pl{s=1}{r} \bigg\{  \f{ G(1 + \nu_{s}+\kappa_s ,1 - \nu_{s}-\kappa_s ) }{    N^{(\nu_{s} + \kappa_s)^2} } \bigg\} \\
\times \pl{a \not= b }{ r } \Big( 1 - \ex{i ( \ov{\bs{t}}_{a-1 } -\ov{\bs{t}}_{b-1}  )}  \Big)^{(\nu_a + \kappa_a)(\nu_b+\kappa_b) }
\cdot  \Big( 1+ \e{o}(1) \Big) \;, 
\end{multline}
where $G$ is the Barnes function. Furthermore, $\bs{\kappa}_r \in \mathbb{Z}^r$ is an integer valued vector 
such that $\bs{\kappa}_r$ maximises
\beq
\bs{\kappa}_r \; \mapsto  \;  -  \sul{s=1}{r} (\nu_{s} + \kappa_s)^2 \; , 
\enq
under the constraint $\ov{\bs{\kappa}}_r=0$. 
Due to the hypothesis on $\nu_a$, $a=1,\dots, r$, the maximizer exists and is unique. 




\subsection{The form factor expansion representation}

In order to get the form factor-like expansion, 
we relabel the integers $\ell_a^{(s)}$ in terms of the particle-hole like integers 
$\{ p_a^{(s)} \}_1^{n^{(s)}}$  and $\{ h_a^{(s)} \}_1^{n^{(s)}}$, for 
some $n^{(s)}=0,\dots, N$. Namely, for any sequence $\ell_1^{(s)} < \dots < \ell_N^{(s)}$ we define the integer 
$n^{(s)}$ and the integers 
\beq
p_1^{(s)} < \dots < p_{n^{(s)}}^{(s)}  \qquad  \e{with} \qquad  p_a^{(s)} \in \mathbb{Z} \setminus \intn{1}{N}  \quad  \e{and} \quad 
 h_1^{(s)} < \dots < h_{n^{(s)}}^{(s)} \quad \e{with} \quad  h_a^{(s)} \in \intn{1}{N} 
\enq
as
\beq
\ell_a^{(s)} \;  = \;  a \qquad \e{for} \;\; a \in \intn{1}{N}\setminus \{h_1^{(s)},\dots, h_{n^{(s)}}^{(s)} \} 
\qquad \e{and} \qquad \ell_{h_a^{(s)}}^{(s)}\; = \; p_a^{(s)} \quad  \e{for} \;\;  a \in \intn{1}{ n^{(s)} }  \;. 
\enq
Then, after some algebra, one gets that 
\bem
\det_N\bigg[ \f{1}{ \la_{\ell_a^{(s-1)}}^{(s-1)} \;  - \;  \la_{\ell_b^{(s)}}^{(s)}  }\bigg] 
\cdot \det_N^{-1}\bigg[ \f{1}{ \la_{a}^{(s-1)} \;  - \;  \la_{b}^{(s)}  }\bigg] \; = \; 
\det_{n^{(s-1)}}\bigg[ \f{ 1 }{ \la_{ h_a^{(s-1)} }^{(s-1)}-\la_{ p_b^{(s-1)} }^{(s-1)} }  \bigg] \cdot 
\det_{n^{(s)}}\bigg[ \f{ 1 }{ \la_{ h_a^{(s)} }^{(s)}-\la_{ p_b^{(s)} }^{(s)} }  \bigg]  \\
\times
\pl{b=1}{n^{(s-1)}} \pl{a=1}{n^{(s)}} 
\bigg\{ \f{  \big( p_b^{(s-1)}-h_a^{(s)} + \nu_s \big) \big( h_b^{(s-1)}-p_a^{(s)} + \nu_s \big)}
{  \big( h_b^{(s-1)}-h_a^{(s)} + \nu_s \big) \big( p_b^{(s-1)}-p_a^{(s)} + \nu_s \big)  } \bigg\} \\
\times \pl{a=1}{n^{(s)}}\bigg\{  \f{ \sin[\pi \nu_s]  }{ \pi } 
\Ga \pab{ N+1-h_a^{(s)}+\nu_s,\;  h_a^{(s)} - \nu_s, \; 1-p_a^{(s)} + \nu_s, \;  N+1-p_a^{(s)} + i0^+  }
{  N+1-h_a^{(s)} , \; h_a^{(s)} , \;  1-p_a^{(s)} + i0^+ , \; N+1-p_a^{(s)} + \nu_s }   \bigg\}  \\
\times \pl{a=1}{n^{(s-1)}}\bigg\{ - \f{ \sin[\pi \nu_{s}]  }{ \pi } 
\Ga \pab{ N+1-h_a^{(s-1)}-\nu_{s},\;  h_a^{(s-1)} + \nu_{s}, \; 1-p_a^{(s-1)}-\nu_{s}, \;  N+1-p_a^{(s-1)} + i0^+  }
{  N+1-h_a^{(s-1)} , \; h_a^{(s-1)} , \;  1-p_a^{(s-1)} + i0^+ , \; N+1-p_a^{(s-1)}-\nu_{s }  } \bigg\} \;. 
\end{multline}
Thus, the series expansion for $\mc{S}_N$ takes the form 
\bem
\mc{S}_N^{(r)} \; =  \; G_N(\{\nu_s\}_1^r; \{t_s\}_1^{r-1} ) \cdot \pl{a=1}{r-1}  \bigg\{ \sul{n^{(s)}=0}{N}  \; 
 \sul{ \substack{ h_1^{(s)}<\dots < h_{n^{(s)}}^{(s)}  \\  h_a^{(s)} \in \intn{1}{N} } }{ }  \; 
 \sul{ \substack{ p_1^{(s)} < \dots < p_{n^{(s)}}^{(s)}  \\  p_a^{(s)} \in \mathbb{Z}\setminus \intn{1}{N} }     }{ }  \bigg\}
 \;\pl{s=1}{r-1}  \mc{H}_N\Big( \{p_a^{(s)}\}_1^{n^{(s)}} \; ; \; \{ h_a^{(s)} \}_1^{n^{(s)}} \mid  \nu_s, \nu_{s+1}; t_s \Big) \\
\times 
\pl{s=1}{r} \pl{b=1}{n^{(s-1)}} \pl{a=1}{n^{(s)}} 
\Bigg\{ \f{  \big( p_b^{(s-1)}-h_a^{(s)} + \nu_s \big) \big( h_b^{(s-1)}-p_a^{(s)} + \nu_s \big)}
{  \big( h_b^{(s-1)}-h_a^{(s)} + \nu_s \big) \big( p_b^{(s-1)}-p_a^{(s)} + \nu_s \big)  } \Bigg\} \, ,
\label{ecriture forme globale somme restreinte FF}
\end{multline}
where
\bem
\mc{H}_N\Big( \{p_a\}_1^n \; ; \; \{ h_a \}_1^n \mid \nu, \eta ; t  \Big) \; = \;  (-1)^n \cdot 
\bigg( \f{\sin [\pi \nu]}{\pi} \cdot \f{\sin [\pi \eta]}{\pi} \bigg)^{n} 
\det_n^2\Big[ \f{1}{p_a-h_b} \Big] \pl{a=1}{n} \Big\{ \ex{it(p_a-h_a)} \Big\} \\
\times \pl{a=1}{n} \Ga \pab{ N+1-h_a+\nu,\;  h_a-\nu, \; 1-p_a+\nu, \;  N+1-p_a + i0^+  }
{  N+1-h_a , \; h_a, \;  1-p_a + i0^+ , \; N+1-p_a + \nu}  \\
\times \pl{a=1}{n}   \Ga \pab{ N+1-h_a-\eta,\;  h_a+\eta, \; 1-p_a-\eta, \;  N+1-p_a + i0^+  }
{  N+1-h_a , \; h_a, \;  1-p_a + i0^+ , \; N+1-p_a-\eta} \;. 
\end{multline}
In \eqref{ecriture forme globale somme restreinte FF}, $G_N(\{\nu_s\}_1^r; \{t_s\}_1^{r-1} ) $ is built out of products of 
so-called background form factor. These have been considered in \cite{KozKitMailSlaTerEffectiveFormFactorsForXXZ}. 
In this free-fermionic setting, each such form factor can be computed explicitly in terms of Barnes functions. 
More explicitly, one has
\bem
G_N(\{\nu_s\}_1^r; \{t_s\}_1^{r-1} ) \; \equiv \; 
\pl{s=1}{r-1} \bigg\{ \ex{-iN t_s \ov{\bs{\nu}}_s}  \bigg\}
 \pl{s=1}{r} \Bigg\{ \bigg( \f{\sin [\pi \nu_s]}{\pi} \bigg)^{N} \det_N\Big[\f{1}{ a-b + \nu_s} \Big]  \Bigg\} \\
\; = \;  \pl{s=1}{r-1} \bigg\{ \ex{-iN t_s \ov{\bs{\nu}}_s}  \bigg\} \cdot 
\pl{s=1}{r} \bigg\{ G\big( 1+\nu_s, 1- \nu_s \big)  \bigg\}   
\cdot  \pl{s=1}{r} \bigg\{  G\pab{ N+1, N+1 }{ N+1-\nu_s, N+1 +\nu_s }   \bigg\}  \\
\; = \;  \f{ \pl{s=1}{r-1} \bigg\{ \ex{-iN t_s \ov{\bs{\nu}}_s}  \bigg\} }{  \pl{s=1}{r} N^{\nu_s^2} }\cdot 
 \pl{s=1}{r-1} \bigg\{ G\big( 1+\nu_s, 1- \nu_s \big)  \bigg\}     \cdot \big( 1+ \e{o}(1)\big) \;. 
\end{multline}

Now we focus in more details on the sum over the integers $\{p_a^{(s)}\}_1^{n^{(s)}}$ and $\{h_a^{(s)}\}_1^{n^{(s)}}$. We will reorganize the sum
into one over excitations of Umklapp type. This will allow us to identify clearly the leading oscillating power,
hence making a comparison with the Fischer--Hartwig asymptotics. 
In the following, we agree upon
\beq
H_L \; = \; \Big\{  1, \dots, \big[ \f{N}{2}\big]  \Big\} \quad , \quad 
H_R \; = \; \Big\{  \big[ \f{N}{2}\big] +1 , \dots, N \Big\} \quad , \quad 
P_L \; =\; -\mathbb{N}^* \quad \e{and} \quad P_R = \mathbb{N} + N \;. 
\enq
Above $[\cdot]$ stands for the floor function. 
It then follows that one can recast the various sums defining $\mc{S}_N^{(r)}$ as
\beq
\mc{S}_N^{(r)}  \; = \; G_N(\{\nu_s\}_1^r; \{t_s\}_1^{r-1} ) 
\pl{s=1}{r-1} \Bigg\{ \sul{n^{(s)}=0}{N} \sul{m^{(s)}=0}{ n^{(s)} }  \sul{r^{(s)}=0}{n^{(s)}}  \bigg\}  
f\Big( \{ m^{(s)}\}_1^{r-1} ;  \{ r^{(s)}\}_1^{r-1} ;  \{ n^{(s)}\}_1^{r-1} \Big)\, ,
\enq
where
\bem
f\Big( \{ m^{(s)}\}_1^{r-1} ;  \{ \ell^{(s)}\}_1^{r-1} ;  \{ n^{(s)}\}_1^{r-1} \Big)\; = \; 
\pl{s=1}{r-1}  \Bigg\{  \sul{ \substack{ h_1^{(s)}<\dots < h_{m^{(s)}}^{(s)}  \\  h_a^{(s)} \in H_L } }{ }  \; 
 \sul{ \substack{ h_{m^{(s)}+1}^{(s)} < \dots < h_{n^{(s)}}^{(s)}  \\  h_a^{(s)} \in H_R } }{ } 
 \sul{ \substack{ p_1^{(s)} < \dots < p_{ r^{(s)} }^{(s)}  \\  p_a^{(s)} \in P_L }     }{ }  \; 
 \sul{ \substack{ p_{r^{(s)} +1 }^{(s)} < \dots  <  p_{ n^{(s)} }^{(s)}  \\  p_a^{(s)} \in P_R }     }{ }   \Bigg\} \\
\times   \pl{s=1}{r-1} 
\bigg\{ \mc{H}_N\Big( \{p_a^{(s)}\}_1^{n^{(s)}} \; ; \; \{ h_a^{(s)} \}_1^{n^{(s)}} \mid  \nu_s, \nu_{s+1}; t_s \Big)  \bigg\}
\cdot \pl{s=1}{r} \pl{b=1}{n^{(s-1)}} \pl{a=1}{n^{(s)}}
 \Bigg\{ \f{  \big( p_b^{(s-1)}-h_a^{(s)} + \nu_s \big) \big( h_b^{(s-1)}-p_a^{(s)} + \nu_s \big)}
{  \big( h_b^{(s-1)}-h_a^{(s)} + \nu_s \big) \big( p_b^{(s-1)}-p_a^{(s)} + \nu_s \big)  } \Bigg\} \; .
\end{multline}

It then remains to make several simplifications in each of these summands by 
using the fact that some terms, due to the largeness of $N$ can be simplified. 
For this purpose, for each set of right or left type integers, we reparameterise the variables in terms of "small" ones by 
\beq
p_a^{(s)}  \; = \;  1-p_{a;-}^{(s)} \;\;\; \e{for} \; \;\; a\, =  \, 1,\dots, r^{(s)} \quad , \quad 
p_a^{(s)} \; = \;  N+ p_{a-r^{(s)};+}^{(s)} \;\;\; \e{for} \; \;\;  a \, = \,  r^{(s)} + 1 ,\dots,  n^{(s)}  \quad , \quad  
\enq
and
\beq
h_a^{(s)} \; = \;  h_{a;-}^{(s)}  \;\;\; \e{for} \; \;\;  a=1,\dots, m^{(s)} \quad , \quad 
h_a^{(s)} \; = \; N-h_{a-m^{(s)};+}^{(s)} \, +\, 1  \;\;\; \e{for} \; \;\;  a=1+ m^{(s)} ,\dots, n^{(s)}   \;. 
\enq

Since we are solely interested in the leading $N\tend +\infty$ asymptotics of each term subordinate to
a choice of $n^{(s)}, r^{(s)}$ and $m^{(s)}$, we shall make several simplifying assumptions. 
First, we shall assume that we can extend the summation up to $\mathbb{N}^{*}$ for all of the 
variables, this without altering the leading $N\tend +\infty$ asymptotics. 
Then, we shall work as if the series were convergent in a strong sense. 
This will allow us to treat the variables $\{p_{a;\pm}^{(s)} \}$ and $\{h_{a ;\pm}^{(s)}\}$
as "small" in respect to $N$. Finally, expanding the summand into powers of $N$,
we shall only keep the leading contribution. 

A long but straightforward calculation then shows that  $\mc{S}_N^{(r)}$ admits the following large-$N$ behaviour 
\bem
\mc{S}_N^{(r)} \; \simeq \;   \pl{s=1}{r}\Big\{ G(1+\nu_s,1-\nu_s) \cdot N^{-\nu_s^2} \Big\} 
\sul{ \substack{ \bs{\ell}_{r-1}  \\ \in \mathbb{Z}^{r-1}  } }{} 
\pl{s=1}{r-1} \bigg\{ \ex{i N t_s (\ell_s- \ov{\bs{\nu}}_s)  }  N^{-2\ell_s^2 }
N^{2(\nu_s-\nu_{s+1})\ell_s} \bigg\}  \pl{s=2}{r-1}\Big\{ N^{2\ell_s \ell_{s-1}}  \Big\} \\
 \times \msc{S}_{\bs{\ell}_{r-1}}^{-}\big( \{t_s\}_1^{r-1}, \{\nu_s\}_1^r \big) \cdot 
   \msc{S}_{\bs{\ell}_{r-1}}^{+}\big( \{t_s\}_1^{r-1}, \{\nu_s\}_1^r \big)  \;. 
\label{ecriture asympt dominantes comme serie part trous}
\end{multline}
Above, the functions $\msc{S}_{\bs{\ell}_{r-1}}^{\pm}$ are as defined in \eqref{definition sommes restreintes type plus et moins}.

\subsection{Identification of the asymptotics and the summation formula}

In order to identify the asymptotics one shall change the variables of summation in
\eqref{ecriture asympt dominantes comme serie part trous} according to 
\eqref{ecriture parametrization kappa1}-\eqref{ecriture parametrization ls et kappa s}. 
Upon using the identities \eqref{ecriture identite resommation exposant oscillant ells vers vars kappas}
\e{and} \eqref{ecriture resomation somme complexe ells vers somme simple ac kappas}, we are, all-in-all, led to 
\beq
\mc{S}_N^{(r)} \; = \;  \pl{s=1}{r-1} \Big\{ \ex{- i N t_s \ov{\bs{\nu}}_s)  }    \Big\} 
\sul{ \substack{ \bs{\kappa}_{r}   \in \mathbb{Z}^{r}  \\ \sum_{s=1}^{r}\kappa_s =0 } }{} 
\pl{s=1}{r} \f{ G(1+\nu_s,1-\nu_s) }{ N^{ (\nu_s+\kappa_s)^2 } }
\pl{s=1}{r-1} \Big\{ \ex{ i N \kappa_{s+1} \ov{\bs{t}}_s  } \Big\}
\cdot 
 \msc{S}_{\bs{\ell}_{r-1}(\bs{\kappa}_r)}^{-}\big( \{t_s\}_1^{r-1}, \{\nu_s\}_1^r \big) \, \cdot  \, 
   \msc{S}_{\bs{\ell}_{r-1}(\bs{\kappa}_r)}^{+}\big( \{t_s\}_1^{r-1}, \{\nu_s\}_1^r \big)  \;. 
\enq
 The identification of the leading asymptotics of the Toeplitz determinant with the leading in $N$ term
of \eqref{ecriture asympt dominantes comme serie part trous} yields
\beq
\msc{S}_{\bs{\ell}_{r-1}(\bs{\kappa}_r)}^{-}\big( \{t_s\}_1^{r-1}, \{\nu_s\}_1^r \big) \cdot 
  \msc{S}_{\bs{\ell}_{r-1}(\bs{\kappa}_r)}^{+}\big( \{t_s\}_1^{r-1}, \{\nu_s\}_1^r \big) 
  \; = \; \pl{s=1}{r}\Big\{ \f{ G(1-\nu_s- \kappa_s) G(1+\nu_s +\kappa_s) }{ G(1-\nu_s) G(1+\nu_s) }  \Big\}
\cdot   \pl{a \not= b }{ r } \Big( 1 - \ex{i ( \ov{\bs{t}}_{a-1 } -\ov{\bs{t}}_{b-1}  )}  \Big)^{(\nu_a + \kappa_a)(\nu_b+\kappa_b) } \;. 
\nonumber
\enq
The parts relative to 
$\msc{S}_{\bs{\ell}_{r-1}(\bs{\kappa}_r)}^{-}\big( \{t_s\}_1^{r-1}, \{\nu_s\}_1^r \big)$ 
(resp. $\msc{S}_{\bs{\ell}_{r-1}(\bs{\kappa}_r)}^{+}\big( \{t_s\}_1^{r-1}, \{\nu_s\}_1^r \big)$)
can be readily inferred by applying a Wiener--Hopf factorization on $\Cx\setminus \R$. 
In order to ensure uniqueness of such a factorization, one should fix the asymptotic behaviour 
of $\msc{S}^{\pm}_{\bs{\ell}_{r-1}(\bs{\kappa}_r)}\big( \{t_s\}_1^{r-1}, \{\nu_s\}_1^r \big)$ at $t \tend \infty$, this non-tangentially to $\R$ and under the constraint $\pm \Im(t) >0$. The latter are readily read-off from the explicit expressions 
\eqref{definition sommes restreintes type plus et moins}
for 
$\msc{S}^{\pm}_{\bs{\ell}_{r-1}(\bs{\kappa}_r)}\big( \{t_s\}_1^{r-1}, \{\nu_s\}_1^r \big)$. Namely
\beq
\msc{S}_{\bs{\ell}_{r-1}(\bs{\kappa}_r)}^{+}\big( \{t_s\}_1^{r-1}, \{\nu_s\}_1^r \big)  \; \underset{t_a \tend + i \infty }{  \sim } \;  
\pl{s=1}{r-1} \Bigg\{  \ex{i  t_s \f{ \ell_s(\ell_s+1) }{2} }  
 G\bigg( \ba{cc}   1+\ell_s-\nu_s , 1+ \ell_s + \nu_{s+1}  \\  
						1-\nu_s , 1+\nu_{s+1} \ea \bigg)   \Bigg\}  \cdot 
\pl{s=2}{r-1} G\bigg( \ba{cc}   1 + \nu_s , 1+ \ell_{s-1}-\ell_s + \nu_s  \\  
						1-\ell_s + \nu_s , 1+\ell_{s-1} + \nu_{s} \ea \bigg)
\nonumber
\enq
and
\beq
\msc{S}_{\bs{\ell}_{r-1}(\bs{\kappa}_r)}^{-}\big( \{t_s\}_1^{r-1}, \{\nu_s\}_1^r \big)  \; \underset{ t_a \tend -i \infty }{  \sim } 
\pl{s=1}{r-1} \Bigg\{  \ex{- i t_s \f{ \ell_s(\ell_s+1) }{2} }  
 G\bigg( \ba{cc}   1-\ell_s+\nu_s , 1- \ell_s - \nu_{s+1}  \\  
						1 + \nu_s , 1 - \nu_{s+1} \ea \bigg)   \Bigg\}  \cdot 
\pl{s=2}{r-1} G\bigg( \ba{cc}   1 - \nu_s , 1 - \ell_{s-1}+\ell_s - \nu_s  \\  
						1-\ell_{s-1} - \nu_s , 1+\ell_{s} - \nu_{s} \ea \bigg)\;. 
\nonumber
\enq
It is then readily checked that one has the decomposition 
\bem
\pl{s=1}{r} \f{ G(1-\nu_s- \kappa_s) G(1+\nu_s +\kappa_s) }{ G(1-\nu_s) G(1+\nu_s) }  \; = \; 
\pl{s=1}{r-1} G\bigg( \ba{cc}   1+\ell_s-\nu_s , 1+ \ell_s + \nu_{s+1}  ,  1-\ell_s+\nu_s , 1- \ell_s - \nu_{s+1}  \\  
						1-\nu_s , 1+\nu_{s+1} , 1 + \nu_s , 1 - \nu_{s+1} \ea \bigg) \\
 \;  \times \; \pl{s=2}{r-1} G\bigg( \ba{cc}   1 + \nu_s , 1+ \ell_{s-1}-\ell_s + \nu_s , 1 - \nu_s , 1 - \ell_{s-1}+\ell_s - \nu_s \\  
						1-\ell_s + \nu_s , 1+\ell_{s-1} + \nu_{s} , 1-\ell_{s-1} - \nu_s , 1+\ell_{s} - \nu_{s} \ea \bigg) \;. 
\end{multline}
The latter was the last missing piece so as to obtain the formulae \eqref{formule calcul exact somme restreinte multiples}. \qed

\end{document}